\documentclass[structabstract]{aa}  
\usepackage{graphicx}
\usepackage{lscape}
\usepackage{natbib}
\usepackage{latexsym}
\newcommand{\HI}{H\,{\sc i}}

\newcommand{\Ha}{H$\alpha$}
\newcommand{\skms}{\ensuremath{\,\mbox{km}\,\mbox{s}^{-1}}}
\newcommand{\kms}{\ensuremath{\mbox{km}\,\mbox{s}^{-1}}}
\newcommand{\vrot}{\ensuremath{v_{\rm rot}}}
\newcommand{\vsys}{\ensuremath{v_{\rm sys}}}

\usepackage{txfonts}
\begin{document}
   \title{Non-circular motions and the cusp-core discrepancy in dwarf galaxies}


   \author{J. van Eymeren
          \inst{1,2,3}
          \and
	  C. Trachternach\inst{2}
	  \and
	  B.~S. Koribalski\inst{3}
	  \and
          R.-J. Dettmar\inst{2}
          }

   \institute{Jodrell Bank Centre for Astrophysics, School of Physics \&
              Astronomy, The University of Manchester, Alan Turing Building,
              Oxford Road, Manchester, M13 9PL, UK\\
              \email{Janine.VanEymeren@manchester.ac.uk}
              \and
              Astronomisches Institut der Ruhr-Universit\"at Bochum,
              Universit\"atsstra{\ss}e 150, 44780 Bochum, Germany
              \and
              Australia Telescope National Facility, CSIRO,
              P.O. Box 76, Epping, NSW 1710, Australia
             }

   \date{Accepted 22 June 2009}

 
  \abstract
   {The cusp-core discrepancy is one of the major problems in astrophysics. It
  results from comparing the observed mass distribution of galaxies with the
  predictions of Cold Dark Matter simulations. The latter predict a cuspy
  density profile in the  inner parts of galaxies, whereas observations of
  dwarf and low surface brightness galaxies show a constant density core.}
   {We want to determine the shape of the dark matter potential in the nuclear
  regions of a sample of six nearby irregular dwarf galaxies.}
   {In order to quantify the amount of non-circular motions which could
  potentially affect a mass decomposition, we first perform a harmonic
  decomposition of the \HI\ Hermite velocity fields of all sample galaxies. We
  then decompose the \HI\ rotation curves into different mass components by
  fitting NFW and pseudo-isothermal halo models to the \HI\ rotation curves
  using a $\chi^2$ minimisation. We model the minimum-disc, the
  minimum-disc\,+\,gas, and the maximum-disc cases.}
   {The non-circular motions are in all cases studied here of the order of
  only a few \kms (generally corresponding to less than 25\%\ of the local
  rotation velocity), which means that they do not significantly affect the
  rotation curves. The observed rotation curves can better be described by
  the cored pseudo-isothermal halo than by the NFW halo. The slopes of the
  dark matter density profiles confirm this result and are in good agreement
  with previous studies. The quality of the fits can often be improved when
  including the baryons, which suggests that they contribute significantly to
  the inner part of the density profile of dwarf galaxies.}
   {}


   \maketitle
%

\section{Introduction}
Early observations of spiral galaxies have shown that their dynamical mass, as
inferred from their surprisingly flat rotation curves, cannot be explained by
luminous matter alone \citep[e.g.,][]{Bosma1978, Rubin1978}. Since then, ``dark
matter'' has become part of modern cosmology. Currently, the most successful
models use a cosmological constant $\rm \Lambda$ as well as collisionless and
dissipationless Cold Dark Matter (CDM). $\rm
\Lambda$CDM simulations like, e.g., the Millennium Simulation by
\citet{Springel2005} have been very successful in describing the observed
large-scale structures in the Universe \citep[][]{Spergel2003,Spergel2007},
but they do not work properly on galaxy scales.

Next to the ``missing satellite'' problem \citep{Moore1999}, the
``cusp-core'' discrepancy \citep[e.g.,][]{deBlok2001b} is still causing a lot
of debate between observers and cosmologists. Numerical simulations
predict cuspy haloes in the inner parts with a density distribution described
by a power law $\rho(r)\sim r^{\alpha}$ with $\alpha$ ranging from -1
\citep[e.g.,][]{Navarro1996} to -1.5
\citep[e.g.,][]{Moore1998,Moore1999}. This cusp leads to a steeply rising
rotation curve. However, observations of dwarf and low surface brightness galaxies show that their rotation curves rise less
steeply than predicted by CDM simulations \citep[e.g.,][]{deBlok2002}. Similar
conclusions have also been made from observations of high surface brightness
disc galaxies \citep{Salucci2001}. At small radii (typically few kpcs), the
mass distribution can better be described by a central, constant-density core
\citep[e.g.,][]{Flores1994,Cote2000,Marchesini2002}.

This interpretation has been met with skepticism by the cosmologists. Thus,
systematic effects in the data like beam smearing, slit misplacement, slit
width, and seeing, as well as the inclination of the galaxy and non-circular
motions have been used to argue against the fact that the observations are
incompatible with cusps
\citep[][]{vandenBosch2000,Swaters2003,Hayashi2004,Spekkens2005}.
\citet{Rhee2004} tried to quantify several of these errors and found that they
add up to 30\%\,-\,50\% underestimation of the rotation curve. However,
repeated one-dimensional long-slit spectra observed by independent observers at
different telescopes \citep{deBlok2003} as well as high-resolution \Ha\
observations using a two-dimensional velocity field
\citep[][]{Simon2005,Kuzio2006,Kuzio2008,Spano2008} rule out
uncertainties related to slit spectroscopy and still suggest the presence of
an isothermal dark matter core in the inner parts of disc
galaxies. Non-circular motions can indeed affect the results of rotation curve
studies as it is generally assumed that the particles are on circular
orbits. Several authors report that CDM haloes are triaxial objects with a
globally elongated potential \citep[e.g.,][]{Hayashi2004,Hayashi2006}. They
suggest that this triaxiality induces large non-circular motions in the inner
parts of galaxies (up to 15\%\ of the maximum rotation velocity), leading to
the observed cored profiles. However, \citet{deBlok2003} simulated rotation
curves showing that non-circular motions of the order of 20\skms\ over a large
fraction of the disc are needed to make them consistent with CDM
haloes. Furthermore, recent observations by \citet{Gentile2005} and
\citet{Trachternach2008} reveal that non-circular motions are typically of the
order of few \kms, i.e., too small to explain the cusp-core discrepancy. This
might not always be the case as \citet{Spekkens2007} found high non-circular
motions in the spiral galaxy NGC\,2976, which have a major effect on the
rotation curve and which they explained by a bar in the inner 500\,pc.

In this paper, the cusp-core discrepancy is addressed by using the rotation
curves derived from \HI\ synthesis data for a mass decomposition of a sample
of six nearby irregular dwarf galaxies. The galaxies have been chosen to be
a subsample of dwarf irregular galaxies in the Local Volume ($D<10$\,Mpc),
which have been observed in \HI\ with sufficiently high spatial resolution
($<1$\,kpc). Some general properties are given in Table~\ref{Genprop}.

In order to rule out systematic effects in the data, we first measure the
non-circular motions. Non-circular motions can have two major causes: chaotic
non-circular motions can be induced for example by star formation
\citep[][]{Oh2008}, systematic non-circular motions relate to
the potential \citep[e.g., spiral arms, tri-axiality of the halo,
  see][]{Schoenmakers1997}. By performing a harmonic decomposition of the
velocity fields, we are able to quantify the systematic non-circular
motions.

This paper is organised as follows: in Sect.~2, the different data sets and
the creation of the velocity fields are briefly described. The velocity fields
and rotation curves are presented in Sect.~3. In Sect.~4, we describe the
procedure of the harmonic decomposition and subsequently show and discuss our
results. Section~5 presents the theoretical background of the mass modelling,
which is followed by a discussion of the results from the mass decomposition
in Sect.~6. In Sect.~7, the main results are summarised.
\begin{table*}
\caption{Some general properties of the sample galaxies.}
\label{Genprop}
$$
\begin{tabular}{lcccc}
\hline
\hline
\noalign{\smallskip}
Galaxy & Type & \multicolumn{2}{c}{optical centre} & \emph{D}\\
& & $\alpha$ (J2000.0) & $\delta$ (J2000.0) & [Mpc]\\
(1) & (2) & (2) & (2) & (3)\\
\hline
\noalign{\smallskip}
NGC\,2366 & IB(s)m & 07$\rm ^h$ 28$\rm ^m$ 54.6$\rm ^s$ & +69\degr\ 12\arcmin\
57\arcsec & 3.44\\
ESO\,059-G001 & IB(s)m & 07$\rm ^h$ 31$\rm ^m$ 18.2$\rm ^s$ & --68\degr\
11\arcmin\ 17\arcsec & 4.57\\
ESO\,215-G?009 & SAB(s)m & 10$\rm ^h$ 57$\rm ^m$ 29.9$\rm ^s$ & --48\degr\
10\arcmin\ 43\arcsec & 5.25\\
NGC\,4861 & SB(s)m & 12$\rm ^h$ 59$\rm ^m$ 02.3$\rm ^s$ & +34\degr\ 51\arcmin\
34\arcsec & 7.50\\
NGC\,5408 &IB(s)m & 14$\rm ^h$ 03$\rm ^m$ 20.9$\rm ^s$ & --41\degr\ 22\arcmin\
40\arcsec & 4.81\\
IC\,5152 & IA(s)m & 22$\rm ^h$ 02$\rm ^m$ 41.5$\rm ^s$ & --51\degr\ 17\arcmin\
47\arcsec & 2.07\\
\noalign{\smallskip}
\hline
\end{tabular}
$$
\footnotesize{Notes: (1) the name of the galaxy; (2) data from NED; (3)
  distance references: NGC\,2366: \citet{Tolstoy1995}, ESO\,059-G001:
  \citet{Karachentsev2006}, ESO\,215-G?009: \citet{Karachentsev2007},
  NGC\,4861: \citet{deVaucouleurs1991}, NGC\,5408: \citet{Karachentsev2002a},
  IC\,5152: \citet{Karachentsev2002b}.}
\end{table*}
\section{Observations and data reduction}
For this study we selected six nearby irregular dwarf galaxies. Some
observational details are given in Table~\ref{HImain}.

\HI\ data for the two northern galaxies, NGC\,2366 and NGC\,4861, were obtained
with the Very Large Array (VLA). The IB(s)m galaxy NGC\,2366 forms part of
``The \HI\ Nearby Galaxy Survey'' \citep[THINGS,][]{Walter2008}. Its peculiar
kinematics  were the subject of several recent publications
\citep[][hereafter dB08, CT08, Oh08, and vE09a]{deBlok2008,
  Trachternach2008,Oh2008,vanEymeren2009a}. The SB(s)m galaxy NGC\,4861 was
studied by \citet{Wilcots1996} and \citet{Thuan2004}. A detailed kinematic
analysis of this galaxy by \citet{vanEymeren2009b} is based on the combined
\HI\ data of two C array and one D array observation. The data reduction
process of NGC\,2366 is described in \citet{Walter2008}, the one of NGC\,4861
in \citet{vanEymeren2009b}.

The four southern dwarf galaxies, ESO\,059-G001, ESO\,215-G?009, NGC\,5408, and
IC\,5152, are selected from the ``Local Volume \HI\ Survey''
\citep[LVHIS\footnote{LVHIS project webpage:
    www.atnf.csiro.au/research/LVHIS/};][]{Koribalski2008,Koribalski2009}. \HI\ data for these galaxies were obtained with the Australia Telescope Compact
  Array (ATCA) and reduced by the LVHIS team \citep[for details
    see][]{Koribalski2009}. All galaxies were observed for
  12 hours each in at least three different configurations (see
  Table~\ref{HImain}).

For the purpose of this paper \HI\ velocity fields are created from the
naturally-weighted data cubes by fitting Gauss-Hermite h3 polynomials to all
line profiles using the GIPSY \footnote{URL:
    http://www.astro.rug.nl/$\sim$gipsy/} \citep[The Groningen Image
  Processing System;][]{vanderHulst1992} task \emph{xgaufit}. This method
allows to accurately define the peak velocities, whereas the standard
intensity-weighted mean velocities are biased toward the longest tail of the
velocity profiles as soon as the distribution is not symmetric (see also
discussion in dB08). In order to separate true emission from noise, we set
some limits for the fitting algorithm (see dB08): the fitted profiles need to
have amplitudes higher than 3$\sigma_{chan}$, where $\sigma_{chan}$ is the
average noise in the line-free velocity channels of the cube. The minimum
dispersion to be fitted has to be higher than the spectral resolution of the
cube (see Table~\ref{HImain} for the noise values and channel
separations). After creating the velocity maps, a small amount of noise pixels
has to be removed that was admitted by the filter criteria. Therefore, we
follow again dB08 and use the integrated \HI\ column density map as a mask:
fits are only retained if the total flux in the integrated \HI\ map is higher
than 3$\sigma_N$ where $\sigma_N$ is the noise in the integrated \HI\ map. It
is defined as $\sqrt{N}\sigma_{chan}$, where \emph{N} is the number of
channels with signal contributing to each pixel. No smoothing is
applied. Altogether, this gives us high quality velocity fields.
\begin{table*}
\caption[The main observational parameters.]{The main observational parameters.}
\label{HImain}
$$
\begin{tabular}{lccccccc}
\hline
  \hline
\noalign{\smallskip}
Galaxy & Telescope & Arrays & Spectral resolution &
  \multicolumn{2}{c}{Synthesised beam} & Noise & References\\
& & & [\kms] & \arcsec\,$\times$ \arcsec & pc\,$\times$ pc & mJy\,beam$^{-1}$&\\
\hline
\noalign{\smallskip}
NGC\,2366 & VLA & B+C+D & 2.6 & 13\,$\times$ 12 & 217\,$\times$ 200 & 0.5 & (1)\\
ESO059-G001 & ATCA & EW352+750C+1.5B & 4 & 51\,$\times$ 47 & 1130\,$\times$
  1041 & 1.3 & (2)\\
ESO215-G?009 & ATCA & EW352+750A+1.5C+6A & 4 & 21\,$\times$ 21 & 530\,$\times$
  530 & 1.0 & (2)\\
NGC\,4861 & VLA & C+D & 5.2 & 31\,$\times$ 30 & 1127\,$\times$ 1091 & 0.6 & (3)\\
NGC\,5408 & ATCA & 375+750D+1.5A & 4 & 57\,$\times$ 50 & 1329\,$\times$ 1166 &
  1.9 & (4)\\
IC\,5152 & ATCA & EW367+750A+1.5A & 4 & 50\,$\times$ 43 & 502\,$\times$ 432 &
1.5 & (2)\\
 \noalign{\smallskip}
\hline
\end{tabular}
$$
\footnotesize{References: (1) \citet{Walter2008}; (2)
  \citet{Koribalski2009}; (3) \citet{vanEymeren2009b}; (4)
  \citet{vanEymeren2008PhD}.}
\end{table*}
\section{The velocity fields and the rotation curves}
\label{Sectvelorot}
\begin{figure*}
\centering
\includegraphics[width=.99\textwidth,bb=34 143 564 793, clip=]{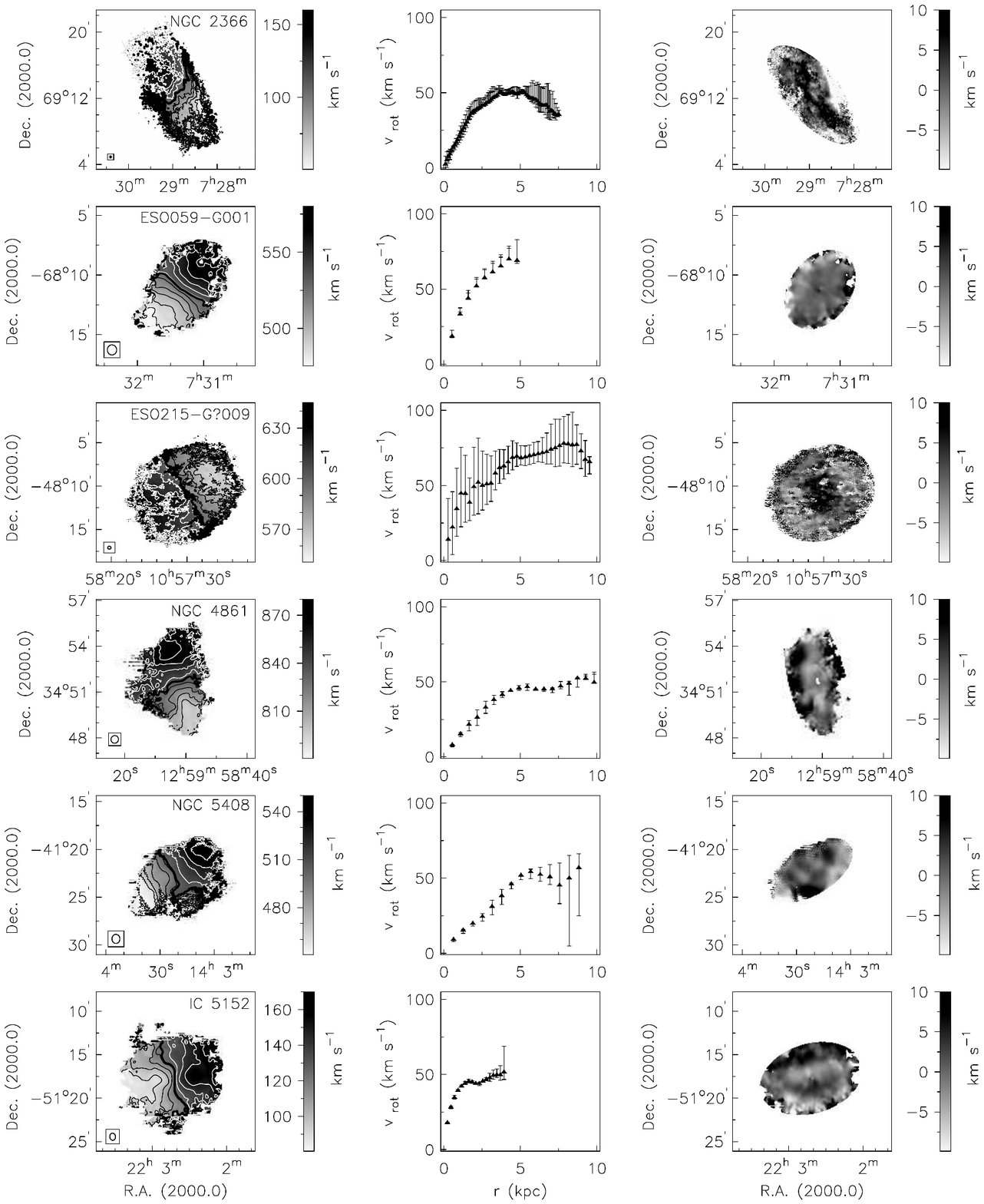}
\caption[]{The \HI\ kinematics of all sample galaxies. {\bf Left panels:}
    Hermite h3 velocity fields. The systemic velocity is marked
    by a bold line, the beam is placed into the lower left corner. Contour
    levels are overlaid in steps of 10\skms. {\bf Middle panels:} the rotation
    curves derived from a tilted-ring analysis (see Table~\ref{Kinpar} for the
    fit parameters). Receding and approaching side are indicated by error
    bars. {\bf Right panels:} the residual maps after subtracting a model
    velocity field (created from the fit parameters in Table~\ref{Kinpar})
    from the observed velocity field.}
\label{velomap_rot}
\end{figure*}

Figure~\ref{velomap_rot} shows the \HI\ velocity fields (left panels). In
order to derive a rotation curve, we use the GIPSY task \emph{rotcur}: tilted rings of half the beam width are fitted to both sides of the \HI\
velocity fields. The systemic velocity, the centre position, the
inclination, and the position angle of each ring are iteratively defined by
keeping all parameters fixed except for the one we want to measure. As initial
estimates we use kinematic parameters derived from fitting isophotes to the
integrated \HI\ intensity map (GIPSY task \emph{ellfit}). The final rotation
curve is then created by keeping the best-fitting parameters fixed. We also
derive rotation curves for receding and approaching side only by keeping the
systemic velocity and the centre position fixed to the values obtained in the
joint approach and by defining the inclination and the position angle
iteratively.

The resulting rotation curves of all sample galaxies are shown in the middle
  panels of Fig.~\ref{velomap_rot}. The error bars of the rotation curves
  indicate the rotation curves of receding (top) and approaching side
  (bottom). We then take the iteratively defined kinematic parameters from the
  tilted-ring analysis (see Table~\ref{Kinpar}) in order to create a model
  velocity field for each galaxy, which is subsequently subtracted from the
  observed velocity field. This shows how well the kinematics of each galaxy
  have been measured in the tilted-ring analysis. The residual maps are
  presented in Fig.~\ref{velomap_rot}, right panels. In most of the cases, the
  derived parameters describe the kinematics of the galaxies quite well and
  the absolute values of the residuals are below an absolute value of
  10\skms. A comparison of the positions of the optical and dynamic centres
  shows that the offsets (given in the last two columns of Table~\ref{Kinpar})
  are generally smaller than one beam size.

A different test to prove the quality of the rotation curves was done, e.g.,
  in vE09a for NGC\,2366. They chose different approaches in order to derive
  the rotation curve. First, the initial estimates from \emph{ellfit} were
  kept fixed. Then, a rotation curve with the best-fitting parameters kept
  fixed was derived (see above). As a third approach, the best-fitting
  parameters were left free. vE09a could show that the deviations between
  these three approaches are generally small with higher deviations in the
  outer parts of the rotation curve. In the inner 1\,kpc, the deviations are
  less than $\pm 2$\skms. We perform this test for all galaxies of our sample
  and generally measure equally small deviations in velocity, at least in the
  inner kpc.

The neutral gas distribution in ESO\,215-G?009 and its kinematics (including
the derivation of a rotation curve) have been studied in detail by
\citet{Warren2004}. Rotation curves of ESO\,059-G001 and ESO\,215-G?009 have
also been derived by \citet{Kirby2009}. A comparison reveals that their
parameters are in good agreement with our values.
\begin{table*}
\caption{Kinematic parameters of the sample galaxies.}
\label{Kinpar}
$$
\begin{tabular}{lcccccccc}
\hline
\hline
\noalign{\smallskip}
Galaxy & \multicolumn{2}{c}{dynamic centre} & \vsys & \vrot & $i$ & $PA$ &
$\alpha_{\rm opt}-\alpha_{\rm dyn}$ & $\delta_{\rm opt}-\delta_{\rm dyn}$ \\
& $\alpha$ (J2000.0) & $\delta$ (J2000.0) & [\kms] & [\kms] & [\degr] &
[\degr] & [\arcsec] & [\arcsec]\\ 
(1) & (2) & (2) & (2) & (2) & (2) & (2) & (3) & (3)\\
\hline
\noalign{\smallskip}
NGC\,2366 & 07$\rm ^h$ 28$\rm ^m$ 53.6$\rm ^s$ & +69\degr\ 12\arcmin\
28\arcsec & 98 & 50 & 63 & 43 & --5 & --29\\
ESO\,059-G001 & 07$\rm ^h$ 31$\rm ^m$ 18.0$\rm ^s$ & --68\degr\ 11\arcmin\
13\arcsec & 528 & 69 & 44 & 325 & --1 & +2\\
ESO\,215-G?009 & 10$\rm ^h$ 57$\rm ^m$ 33.0$\rm ^s$ & --48\degr\ 10\arcmin\
47\arcsec & 599 & 77 & 28 & 119 & +31 & --4\\
NGC\,4861 & 12$\rm ^h$ 59$\rm ^m$ 01.4$\rm ^s$ & +34\degr\ 51\arcmin\
44\arcsec & 835 & 46 & 67 & 15 & --11 & +10\\
NGC\,5408 & 14$\rm ^h$ 03$\rm ^m$ 22.7$\rm ^s$ & --41\degr\ 22\arcmin\
12\arcsec & 502 & 53 & 58 & 300 & +20 & +28\\
IC\,5152 & 22$\rm ^h$ 02$\rm ^m$ 41.9$\rm ^s$ & --51\degr\ 17\arcmin\
45\arcsec & 123 & 44 & 52 & 284 & +4 & +2\\
\noalign{\smallskip}
\hline
\end{tabular}
$$
\footnotesize{Notes: (1) the name of the galaxy; (2) kinematic parameters
  derived by fitting a tilted-ring model to the \HI\ Hermite velocity fields,
  the position angle is measured counter-clockwise from north to the receding
  side of the galaxies; (3) position offset of the dynamical centre in
  comparison to the optical centre.}
\end{table*}
\section{Harmonic decomposition}
\label{Sectharmdecomp}
In order to search for non-circular motions, we decompose the velocities
detected along the tilted rings into multiple terms of sine and
cosine. Following \citet{Schoenmakers1999PhD}, the line of sight velocity
$v_{\rm los}$ can be described as:
\begin{equation}
v_{\rm los}(r)=v_{\rm sys}(r)+\sum\limits_{m=1}^{N}{c_{\rm
    m}(r)\cos{m\,\psi}+s_{\rm m}(r)\,\sin{m\,\psi}},
\label{eqlineofsight}
\end{equation}
where \emph{N} is the maximum fit order used, \emph{r} is the radial distance
from the dynamic centre, $\psi$ is the azimuthal angle in the plane of the
disc, and $v_{\rm sys}$ is the 0th order harmonic component $c_0$. As shown in
CT08, a decomposition of the velocity field up to third order is sufficient to
capture most of the non-circular signal. Therefore, we restrict our analysis
to $N=3$. In case of purely circular motion, only $m=0$ and $m=1$ terms are
included into Eq.~\ref{eqlineofsight}.

The decomposition is performed with the GIPSY task \emph{reswri}, which fits
a tilted-ring model to the original velocity field using circular rotation,
and decomposes the line-of-sight velocity along each ring into multiple terms
of sine and cosine. Subsequently, a model velocity field is created and
subtracted from the original velocity field producing a residual velocity
field. Note that the centre coordinates should be kept fixed
\citep[see][]{Schoenmakers1999PhD}. We run the routine with two different
parameter sets: in the first case, all parameters except for the centre
position (listed in Table~\ref{Kinpar}) are left free. In the second case,
inclination and position angle are fixed to the values given in
Table~\ref{Kinpar}.

The quadratically-added amplitude for each order of the harmonic decomposition
is calculated using
\begin{equation}
A_1(r)=\sqrt{s^2_1(r)}
\label{eqAmp1}
\end{equation}
for $m=1$ ($c_1$ is the circular rotation velocity) and
\begin{equation}
A_{\rm m}(r)=\sqrt{c^2_{\rm m}(r)+s^2_{\rm m}(r)}
\label{eqAmpm}
\end{equation}
for $m>1$. Additionally, the quadratically-added amplitude of all non-circular
harmonic components (up to $N=3$ in this case) was calculated from
\begin{equation}
A_{\rm r}(r)=\sqrt{s^2_1(r)+c^2_2(r)+s^2_2(r)+c^2_3(r)+s^2_3(r)}.
\label{eqAmpr}
\end{equation}
We also derive the median value of $A_m(r)$ for each harmonic order $m$ in
two different ways: once, using the entire radial range and once, using the
inner 1\,kpc only, which is the region where the distinction between a cusp
and a core becomes most obvious \citep{deBlok2004}. More information about
harmonic decompositions in general can be found in \citet{Schoenmakers1997}.

In the following subsections the results for each galaxy are discussed
separately. For a better understanding of the interpretation see the rules of
thumb given in \citet{Schoenmakers1999PhD}.
\subsection{NGC\,2366}
\label{SectN2366harm}
\begin{figure*}
 \centering
\includegraphics[width=\textwidth,viewport=48 173 570 653, clip=]{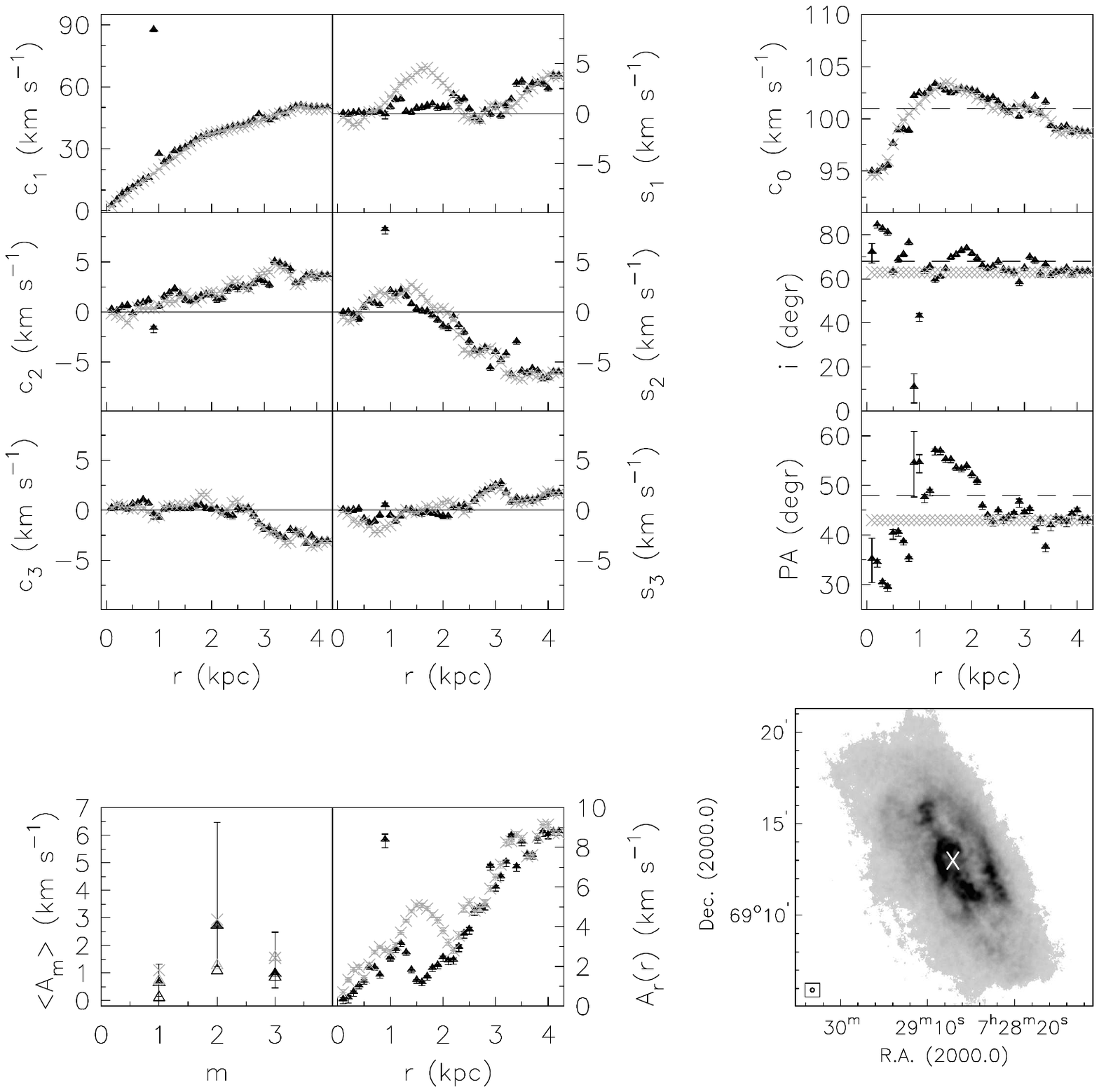}
\caption[]{The results of the unconstrained (black triangles) and constrained
  (light grey crosses) harmonic decomposition of NGC 2366. For the constrained
  fit the inclination and the position angle were fixed to the values given in
  Table~\ref{Kinpar}. {\bf Upper left:} following Eq.~\ref{eqlineofsight},
  circular ($c_1$) and non-circular ($c_2$, $c_3$, $s_1$, $s_2$, $s_3$)
  harmonic components corrected for inclination and plotted \emph{vs.} radius
  \emph{r}. {\bf Upper right:} systemic velocity \emph{c}$_0$, inclination
  \emph{i}, and position angle \emph{PA}, again plotted \emph{vs.} radius
  \emph{r}. The dashed horizontal lines represent the error weighted means of
  the unconstrained fit. The inclination and position angle are from the
  tilted-ring fit assuming circular rotation. For the unconstrained case the
  error bars represent the formal errors. {\bf Lower left:} the median
  amplitudes of the individual harmonic components derived from
  Eqs.~\ref{eqAmp1} and \ref{eqAmpm} plotted \emph{vs.} harmonic number
  \emph{m}. The error bars denote the upper and lower quartile of the
  distribution of the unconstrained $<A_{\rm m}>$. The open triangles
  represent the median amplitudes within the inner 1\,kpc. {\bf Lower middle:}
  $A_{\rm r}(r)$, the quadratically-added amplitude of all non-circular
  components derived from Eq.~\ref{eqAmpr}. The uncertainties are estimated
  using Gaussian error propagation. {\bf Lower right:} the \HI\ intensity
  distribution. The dynamic centre is marked by a white cross.}
\label{N2366harm}
\end{figure*}

Using accurately defined centre coordinates, CT08 have already performed a
harmonic decomposition of the Hermite velocity field of NGC\,2366 (see also
dB08). We also use the Hermite velocity field, but different centre
coordinates (see Table~\ref{Kinpar}). The radial
distribution of all fitted parameters is shown in Fig.~\ref{N2366harm}. The
black filled triangles represent the case where all parameters except for the
centre position are left free. The inclination varies significantly in the
inner 1.5\,kpc. As NGC\,2366 is dominated by close-to solid body rotation, it
is difficult to determine \vrot\ and \emph{i} simultaneously. At a radius of
1\,kpc, $c_0$ rises from 95 to 102\skms\ and the position angle jumps from
about 35\degr\ to 55\degr. At the same radius, the inclination has very low
values, which leads to very high values of the inclination corrected
$c_1$. This is probably caused by the sudden decrease in \HI\ intensity. At a
radius of about 3\,kpc, the $s_1$ and $s_3$ terms show the characteristic
wiggles caused by spiral arms. As discussed by \citet{Tikhonov2008} and vE09a,
NGC\,2366 shows evidence for two weak spiral arms.

The light grey crosses in Fig.~\ref{N2366harm} show the constrained case where
we use a fixed inclination of 63\degr\ and a fixed position angle of 43\degr\
(see Table~\ref{Kinpar}). For most of the tilted rings the values are in very
good agreement with the ones of the unconstrained fit. A pronounced deviation
can be seen in the $s_1$ term where the constrained fit leads to higher
amplitudes between 1 and 2\,kpc than the unconstrained fit. This is due to the
constant position angle which differs significantly from the unconstrained
values.

The median value of the quadratically-added amplitude for each order is given
in the lower left panel of Fig.~\ref{N2366harm}. In both cases the amplitudes
are below 3\skms. As the innermost region of a
galaxy is most significant for the cusp-core debate \citep{deBlok2004}, we
also show the median values within 1\,kpc (open triangles). These are
generally below 1\skms\ and therefore even smaller than the ones averaged over
the entire radial range. $A_{\rm r}(r)$ is below 3\skms\ in the inner 2.5\,kpc
and rises to 9\skms\ in the outer parts contributing about 9\%\ (constrained
case) and 6\%\ (unconstrained case) to the maximum rotation velocity of
NGC\,2366 (see Table~\ref{tabharm}).

In general, our results agree very well with the results of CT08 despite the
different centre coordinates. This is to be expected as CT08 have already
addressed the problem of inaccurate centre positions. They showed that for
dwarf galaxies, which are characterised by solid body rotation in the inner
parts, the results of the harmonic decomposition are insensitive to small
offsets in the centre position.
\subsection{ESO\,059-G001}
\begin{figure*}
 \centering
\includegraphics[width=\textwidth,viewport=48 173 570 653, clip=]{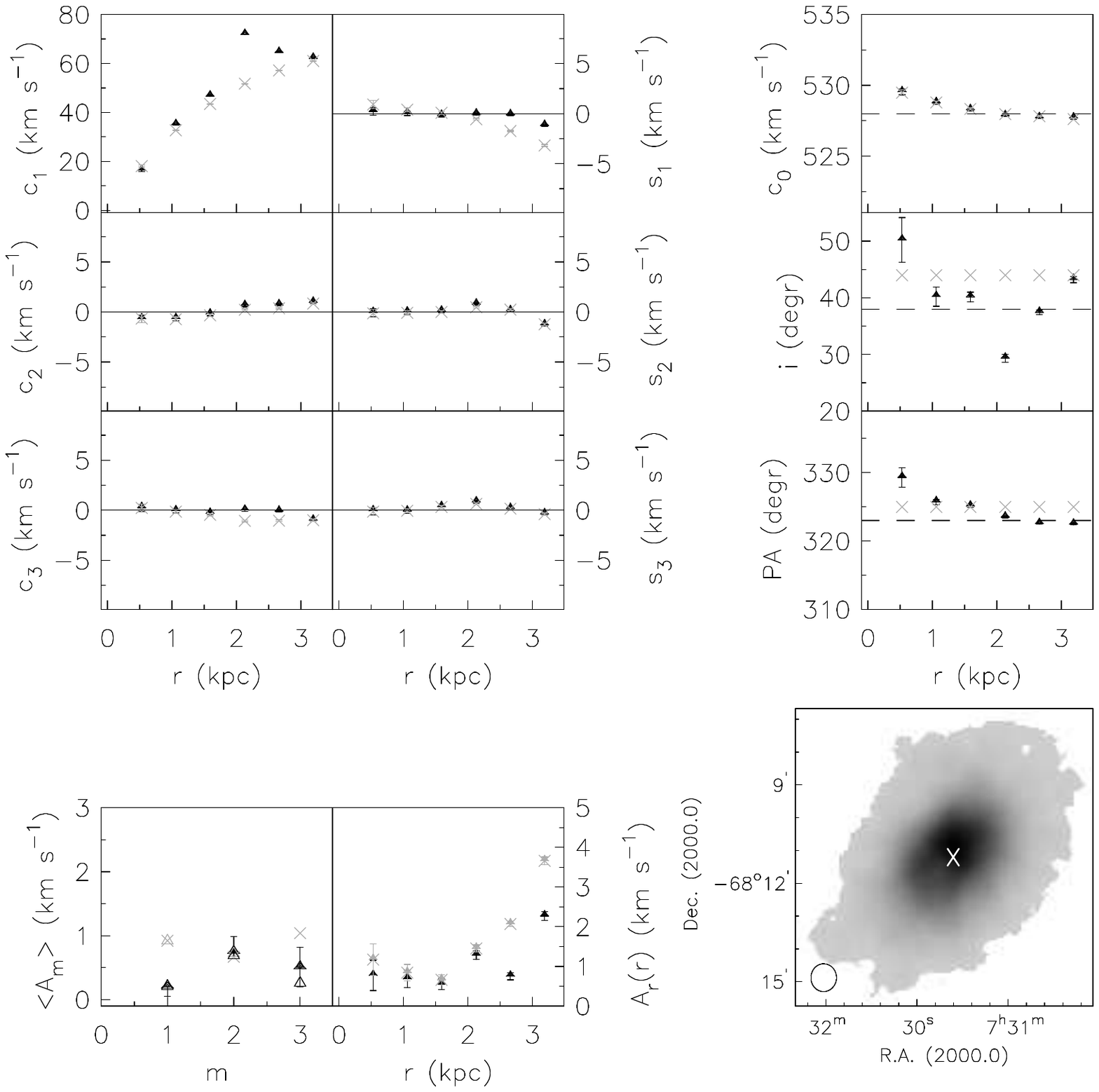}
\caption[]{The same as in Fig.~\ref{N2366harm} for ESO\,059-G001. The
  inclination and the position angle for the constrained fit are listed in
  Table~\ref{Kinpar}.}
\label{eso059-g001harm}
\end{figure*}

The results of the harmonic decomposition of ESO\,059-G001 are shown in
Fig.~\ref{eso059-g001harm}. The inclination at 2\,kpc drops to 30\degr, which
leads to deviations of the rotation velocity. The position angle varies only
slightly over the entire radial range. The amplitudes of the non-circular
components are the smallest of all galaxies studied here. \citet{Parodi2002}
report ``rudiments of spiral arms'' in ESO\,059-G001, which can, however, not
be seen in the harmonic components (probably due to their weakness and the
relatively low spatial resolution of the \HI\ data).

The median amplitudes are below 1\skms. $A_{\rm r}(r)$ stays below 3\skms\
over the entire radial range and only contributes 2\%\ (constrained case) and
1\%\ (unconstrained case) to the maximum rotation velocity (see
Table~\ref{tabharm}).

ESO\,059-G001 is dominated by close-to solid body rotation and additionally,
the inclination is low. A tilted-ring analysis with free parameters is
therefore difficult to perform. However, the constrained case with a fixed
inclination of 44\degr\ and a position angle of 325\degr\ still gives
acceptable values. As long as \emph{i} does not differ too much from the
median value, the fits are in good agreement with the unconstrained ones. The
residual map given in Fig.~\ref{velomap_rot} also shows that the kinematic
parameters derived from the constrained approach represent the galaxy quite
well. 
\subsection{ESO\,215-G?009}
\label{SectESO215-G009harm}
\begin{figure*}
 \centering
\includegraphics[width=\textwidth,viewport=48 173 570 653, clip=]{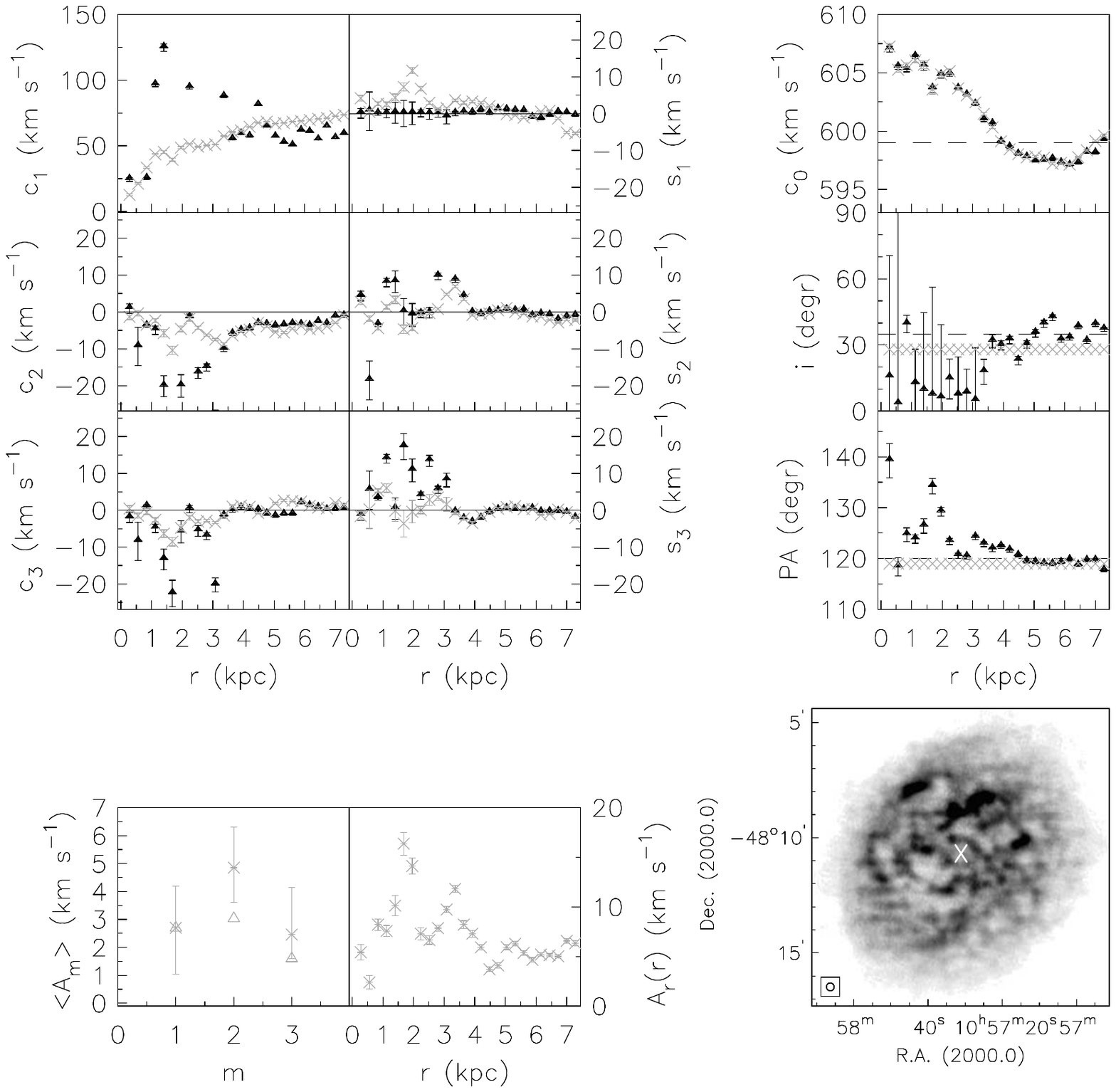}
\caption[]{The same as in Fig.~\ref{N2366harm} for ESO\,215-G?009. The
  inclination and the position angle for the constrained fit are listed in
  Table~\ref{Kinpar}. The analysis of this galaxy is based on the constrained
  case (see Sect.~\ref{SectESO215-G009harm}). Therefore, no $<A_{\rm m}>$ and
  $A_{\rm r}(r)$ are given for the unconstrained case.}
\label{eso215-g009harm}
\end{figure*}

ESO\,215-G?009 has a strongly varying and very low inclination in the inner
3.5\,kpc (see Fig.~\ref{eso215-g009harm}). As shown for ESO\,059-G001, a low
inclination makes it very difficult to perform a tilted-ring analysis by
leaving all parameters free. The partially unreasonable values for the
inclination in the inner 3.5\,kpc of the unconstrained case (with \emph{i} as
low as 5\degr) lead to an unphysical rotation curve. A constrained fit with
fixed inclination and position angle (see Table~\ref{Kinpar}) gives much more
reasonable results which are, apart from the inner 3.5\,kpc, largely
consistent with the results from the unconstrained case. Therefore, we decide
to ignore the unconstrained case for the subsequent analysis of ESO\,215-G?009
and to use the constrained case instead.

Most of the amplitudes of
the harmonic components are close to zero. The values of $A_{\rm r}(r)$ are
typically below 10\skms\ contributing about 8.5\%\ to the maximum rotation
velocity (see Table~\ref{tabharm}). The median amplitudes are below 5\skms, in
the inner kpc below 3\skms.

We also show the results of the unconstrained fit: the low inclination causes
very large circular velocity components (note that we do not plot the values
between 1 and 3\,kpc that have high formal errors). The low inclination
becomes also noticeable in the non-circular components, which are partly as
high as 20\skms.
\subsection{NGC\,4861}
\begin{figure*}
 \centering
\includegraphics[width=\textwidth,viewport=48 173 570 653, clip=]{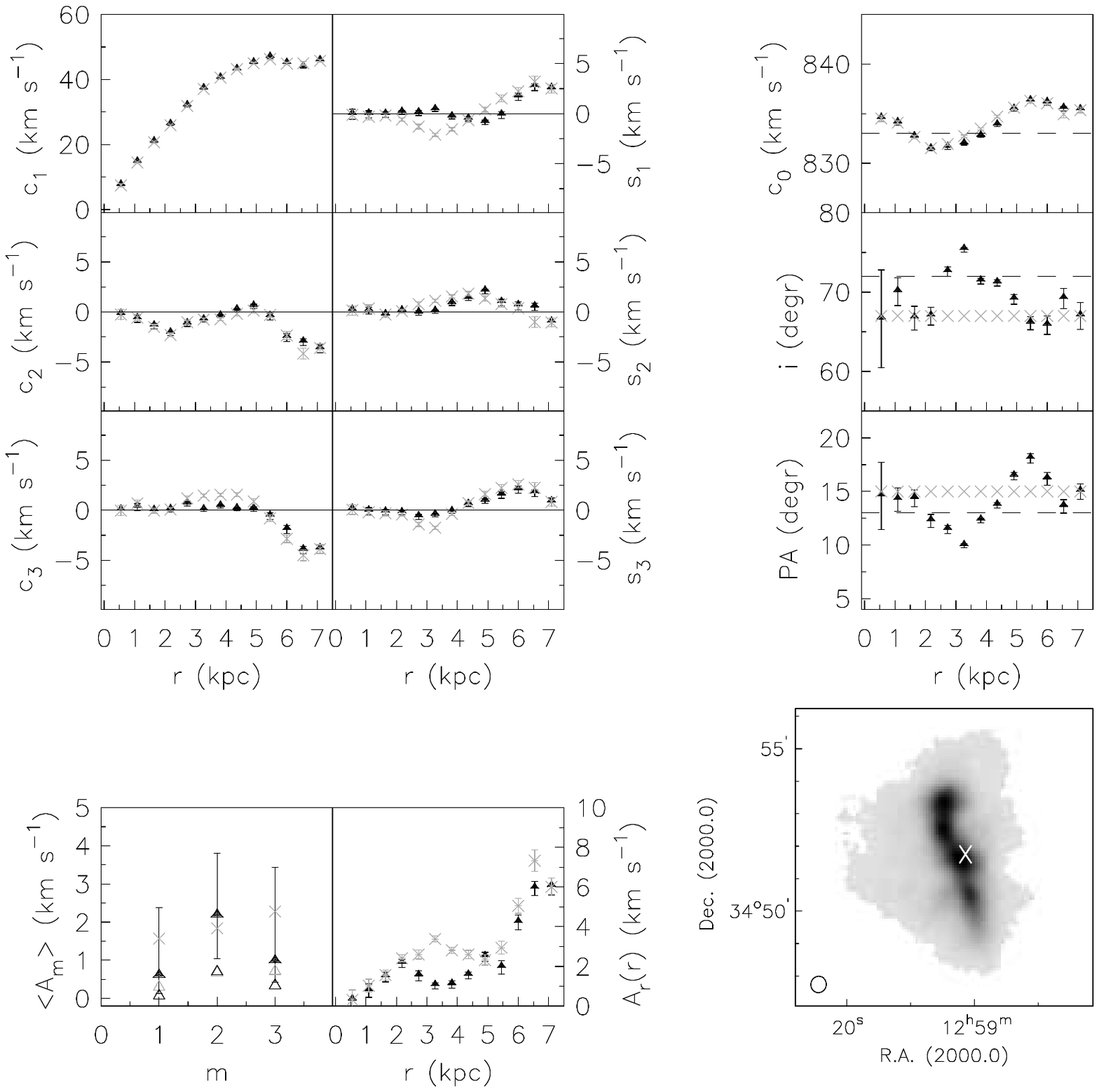}
\caption[]{The same as in Fig.~\ref{N2366harm} for NGC\,4861. The inclination
  and the position angle for the constrained fit are listed in
  Table~\ref{Kinpar}.}
\label{N4861harm}
\end{figure*}

The results of the harmonic decomposition of NGC\,4861 are given in
Fig.~\ref{N4861harm}. The amplitudes of the non-circular components are
usually close to zero. No pronounced deviation can be seen, which fits into
the picture of a quiescent galaxy without spiral arm structure as described in
\citet{vanEymeren2009b}. The median values of the quadratically-added
amplitudes for each harmonic order are below 2.5\skms\ over the entire radial
range and even below 1\skms\ in the inner 1\,kpc. The distribution of $A_{\rm
  r}(r)$ shows that the amplitude is generally below 3\skms, except for the
outermost part. On average, the non-circular motions contribute about 6\%\
(constrained case) and 3\%\ (unconstrained case) to the maximum rotation
velocity. The amplitudes of the harmonic components of the constrained fit
(inclination of 67\degr, position angle of 15\degr) usually agree to within
2\skms\ with the ones from the unconstrained fit.
\subsection{NGC\,5408}
\begin{figure*}
 \centering
\includegraphics[width=\textwidth,viewport=48 173 570 653, clip=]{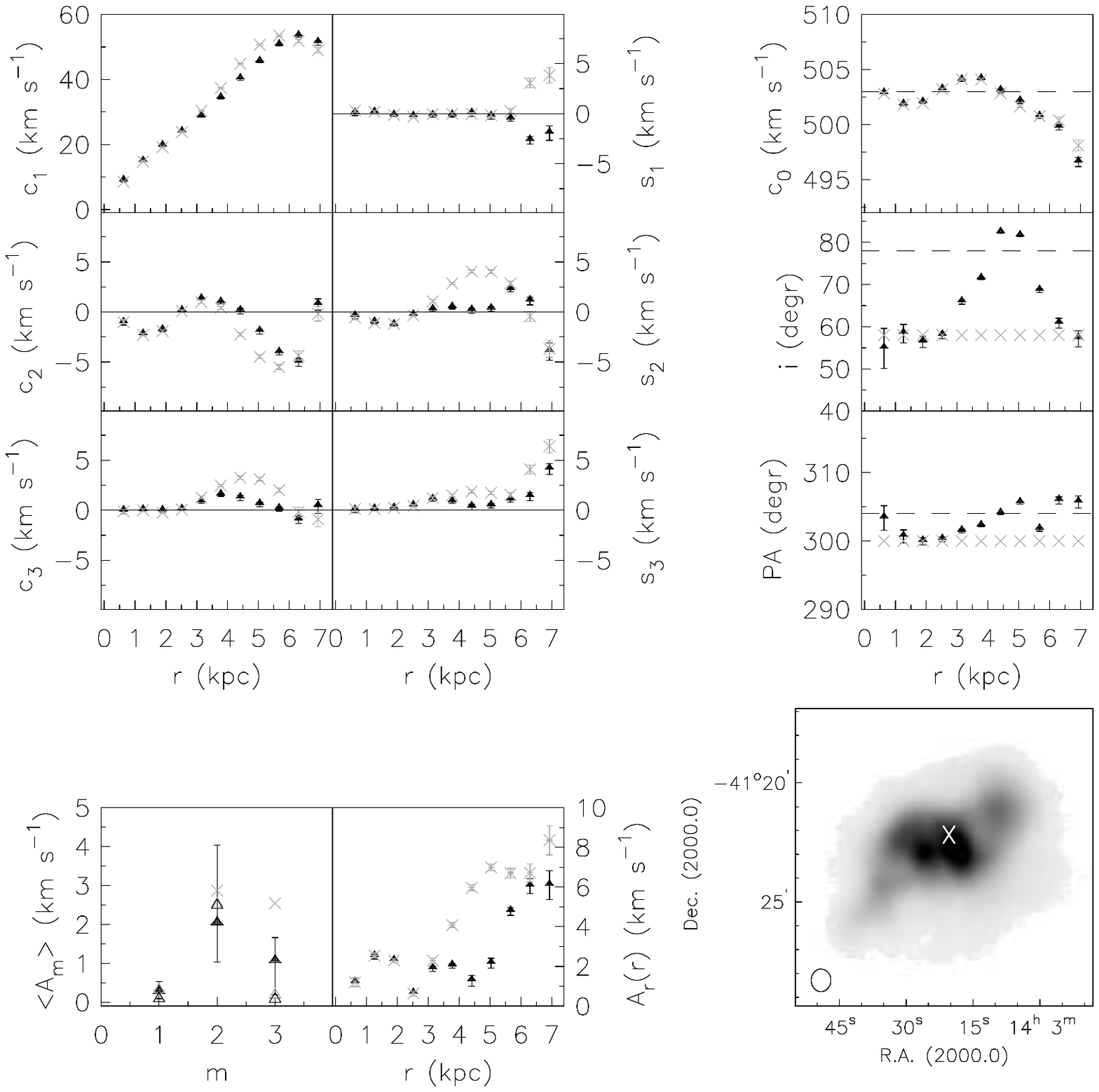}
\caption[]{The same as in Fig.~\ref{N2366harm} for NGC\,5408. The inclination
  and the position angle for the constrained fit are listed in
  Table~\ref{Kinpar}.}
\label{N5408harm}
\end{figure*}

Figure~\ref{N5408harm} shows the results for NGC\,5408. In her PhD thesis,
\citet{vanEymeren2008PhD} suggested that this galaxy has two
kinematic systems, which is probably the reason for the clear and well-defined
radial variation of the inclination in the unconstrained fit. Nevertheless,
the amplitudes of the non-circular motions are small. $A_{\rm r}(r)$ is below
3\skms\ in the inner 5\,kpc and rises to about 6\skms\ in the outer parts. On
average, the non-circular motions contribute less than 8\%\ (constrained case)
and 4\%\ (unconstrained case) to the maximum
rotation velocity. The median amplitudes of the individual harmonic orders are
below 3\skms\ independent of the radial range we look at.

For the constrained case we use a fixed inclination of 58\degr\ and a fixed
position angle of 300\degr\ (see Table~\ref{Kinpar}). In the inner 2.5\,kpc
this inclination agrees with the one from the unconstrained fit. However, it
differs significantly from the inclination of the outer kinematic system. For
$r>2.5$\,kpc this results in different inclination corrections for the
constrained and the unconstrained cases, which can be seen, e.g., in the small
offset in $c_1$.
\subsection{IC\,5152}
\label{SectIC5152harm}
\begin{figure*}
 \centering
\includegraphics[width=\textwidth,viewport=48 173 570 653, clip=]{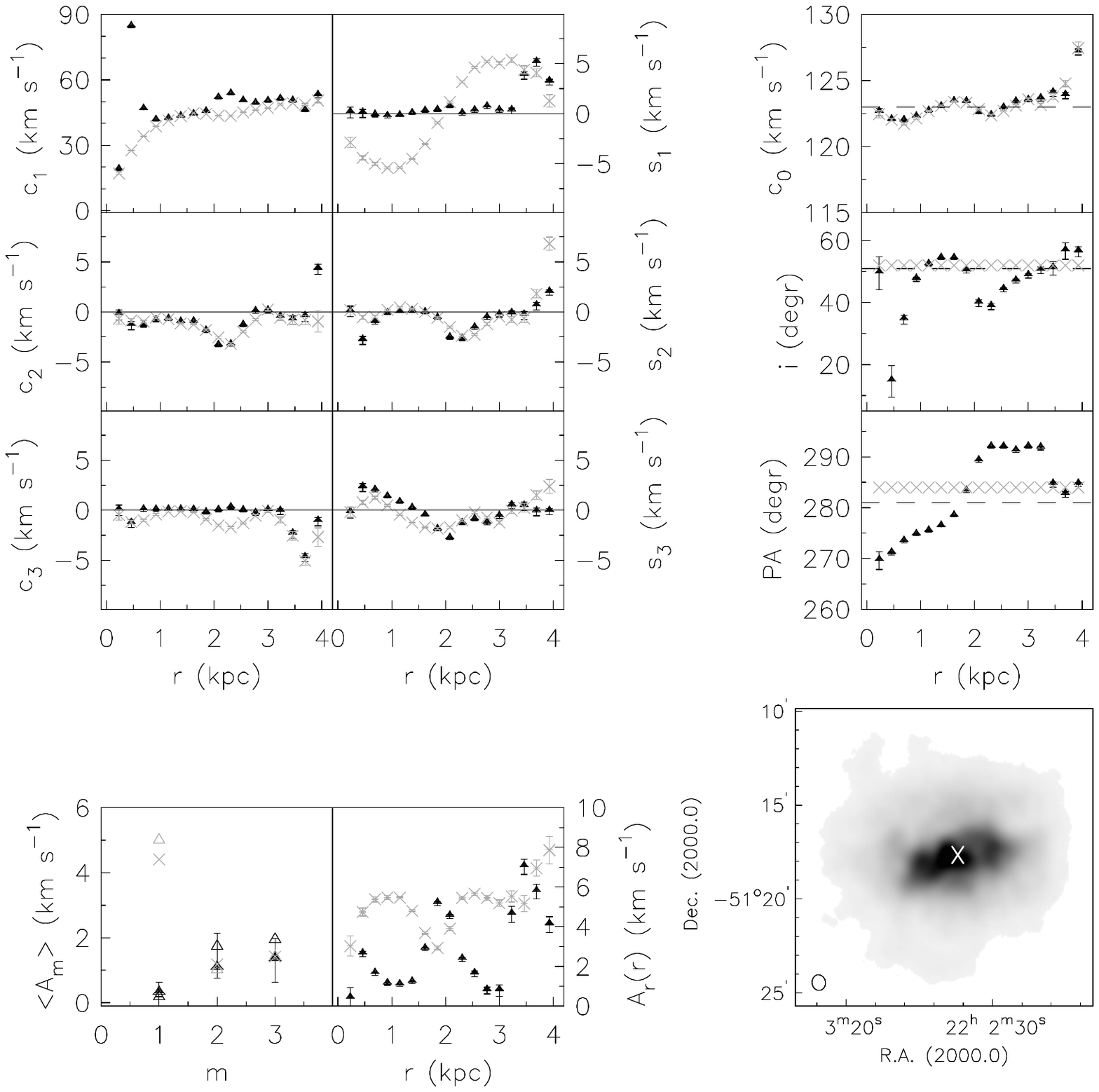}
\caption[]{The same as in Fig.~\ref{N2366harm} for IC\,5152. The inclination
  and the position angle for the constrained fit are listed in
  Table~\ref{Kinpar}.}
\label{IC5152harm}
\end{figure*}

Except for one outlier in the inner 1\,kpc, the inclination of IC\,5152 is
roughly constant with radius (see Fig.~\ref{IC5152harm}). The position angle
increases at 2\,kpc by about 20\degr. Altogether, this leads to small
non-circular components, often close to 0\skms. The constrained case (with an
inclination of 45\degr\ and a position angle of 287\degr) and the
unconstrained case are in good agreement, although the fixed position angle
leads to higher amplitudes in the $s_1$ term. $A_{\rm r}(r)$ varies over the entire radial
range, but stays below 8\skms. The difference between the amplitudes of the
constrained and unconstrained values are several \kms, which is due to the
difference in the $s_1$ term. The median amplitudes of the individual
harmonic orders are below 2\skms\ for the unconstrained case. In order to
calculate the median amplitudes in the inner 1\,kpc, we removed the outlier at
0.5\,kpc. Especially the $m=1$ term of the constrained case is high, which is
again due to the high amplitudes of the $s_1$ term. In comparison to the other
galaxies of our sample, IC\,5152 shows the highest contribution of
non-circular motions to the maximum rotation velocity.
\subsection{Discussion}
In all sample galaxies, the non-circular motions contribute on average about
8\%\ (constrained case) and about 4\%\ (unconstrained case) to the maximum
rotation velocity. However, in the inner few kpcs the rotation velocity has
not reached its maximum value, yet. Furthermore, for the cusp-core debate the
inner few kpcs are most relevant, whereas non-circular motions in the outer
parts of a galaxy, though surely interesting, are of no concern. Therefore, we
perform a more detailed analysis of the quadratically-added amplitude of all
non-circular motions within the inner 5\,kpc. We concentrate on the results of
the constrained case. Especially in the inner few kpcs, a strong change in
inclination or position angle is not to be expected so that it is a reasonable
assumption to keep both parameters fixed. As Table~\ref{tabharm} shows, the
amplitudes of the constrained case are usually up to a factor of two higher
than the amplitudes of the unconstrained case. Therefore, the constrained case
gives some kind of upper limit to the amount of non-circular motions.

\begin{figure*}
 \centering
\includegraphics[width=\textwidth,viewport=63 429 565 663, clip=]{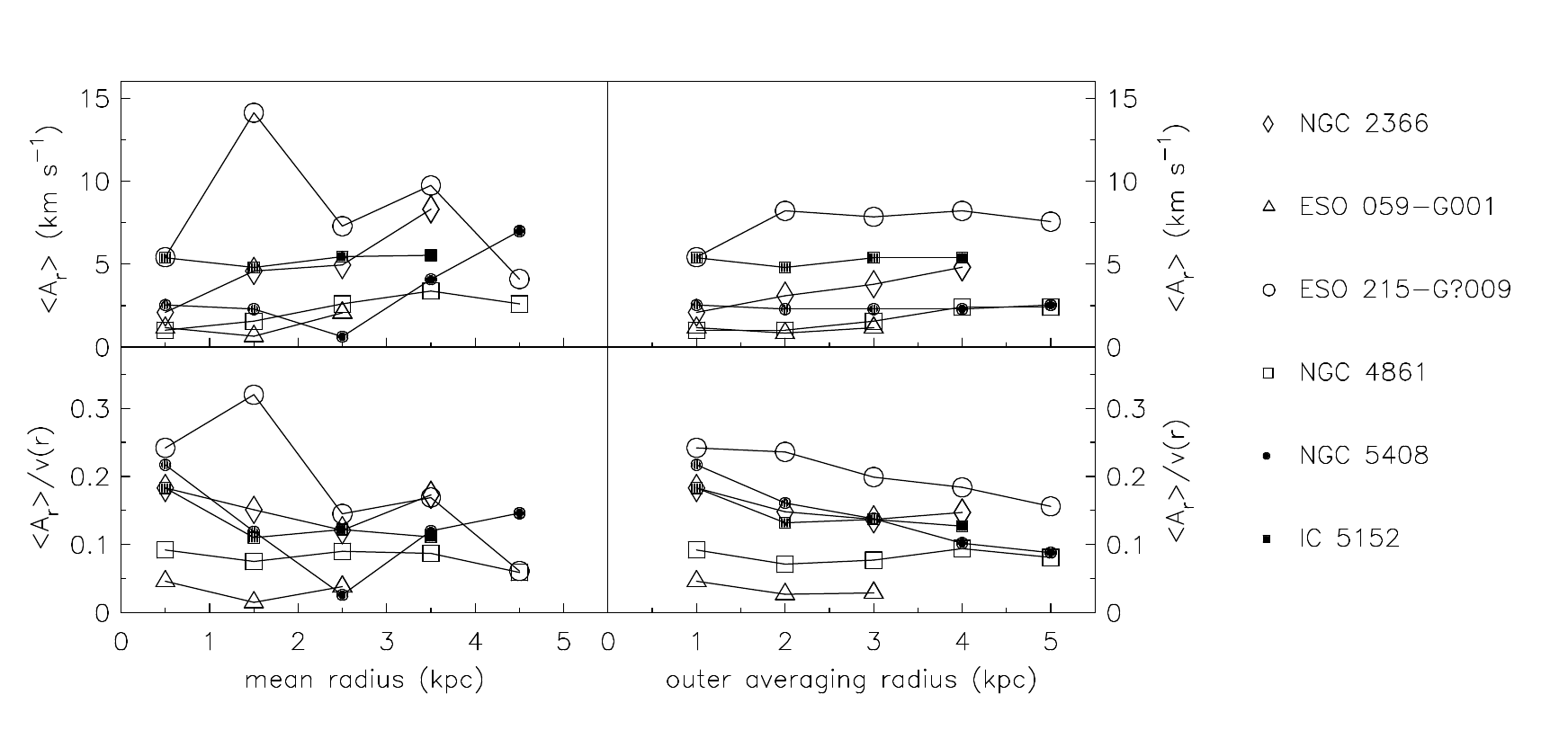}
\caption[]{The mean of the quadratically-added amplitudes derived from the
  constrained case. {\bf Upper left panel:} the amplitudes of the non-circular
  motions within rings of 1\,kpc width (i.e., $0<r<1$\,kpc, $1<r<2$\,kpc, ...,
  $4<r<5$\,kpc) for each galaxy (indicated by different symbols). {\bf Upper
  right panel:} like the upper left panel, but the amplitudes of the
  non-circular motions are averaged within rings of increasing radius (i.e.,
  $0<r<1$\,kpc, $0<r<2$\,kpc, ..., $0<r<5$\,kpc). {\bf Lower left and right
  panel:} like the upper left and right panel, but the amplitudes of the
  non-circular motions are normalised by the local rotation velocity.}
\label{ampvsr}
\end{figure*}

Figure~\ref{ampvsr} shows the amplitudes of the constrained case averaged
within rings of 1\,kpc width (left panels) and within rings of increasing
radius (right panels). Furthermore, we differentiate between the absolute
amplitudes (upper panels) and the amplitudes which are normalised by the local
rotation velocity (lower panels). In all sample galaxies and for all radii, the
non-circular motions generally contribute less than 25\,\%\ to the local
rotation velocity, sometimes even less than 10\,\%. In the inner 1\,kpc,
$<A_{\rm r}(r)>$ is often close to 1 or 2\skms\ (see Table~\ref{tabharm}),
which corresponds to about 5 to 25\%\ of the local rotation
velocity. According to the simulations by \citet{Hayashi2004} and
\citet{Hayashi2006}, the non-circular motions add up to about 50\%\ of the
local rotation velocity at a radius of 1\,kpc and even more below 1\,kpc.
\begin{table*}
\caption[]{Derived quantities from the harmonic decomposition.}
\label{tabharm}
$$
\begin{tabular}{lccccccccc}
\hline
  \hline
\noalign{\smallskip}
Galaxy & $<A_{\rm r}>$ & $<A_{\rm r,1\,kpc}>$ & $<A_{\rm r}>$/$v_{\rm rot}$ & $<M_{\rm res}>$ & $<A_{\rm r,c}>$ & $<A_{\rm r,c,1\,kpc}>$ & $<A_{\rm r,c}>$/$v_{\rm rot}$ & $<M_{\rm res,c}>$ & $r_{\rm max}$\\
& [\kms] & [\kms] & [\%] & [\kms] & [\kms] & [\%] & [\kms] & [kpc]\\
(1) & (2) & (3) & (4) & (5) & (6) & (7) & (8) & (9) & (10)\\
\hline
\noalign{\smallskip}
NGC\,2366 & $2.91^{+4.09}_{-1.06}$ & $1.50^{+0.42}_{-0.87}$ & 5.8 & 2.15 &
$4.81^{+2.32}_{-1.83}$ & $2.08^{+0.66}_{-0.67}$ & 9.6 & 1.89 & 4.1\\
\noalign{\smallskip}
ESO\,059-G001 & $0.77^{+0.51}_{-0.09}$ & $0.77^{...}_{...}$ & 1.1 & 0.43 &
$1.45^{+0.62}_{-0.61}$ & $1.17^{...}_{...}$ & 2.1 & 0.37 & 3.2\\
\noalign{\smallskip}
ESO\,215-G?009 & ... & ... & ... & ... & $6.57^{+1.64}_{-1.38}$ & $5.42^{+2.79}_{-3.03}$ & 8.5 & 1.99 & 7.3\\
\noalign{\smallskip}
NGC\,4861 & $1.54^{+0.94}_{-0.46}$ & $0.78^{...}_{...}$ & 3.3 & 1.04 &
$2.60^{+0.78}_{-0.28}$ & $1.02^{...}_{...}$ & 5.7 & 1.07 & 7.1\\
\noalign{\smallskip}
\noalign{\smallskip}
NGC\,5408 & $2.15^{+2.62}_{-0.88}$ & $2.50^{...}_{...}$ & 4.1 & 1.92 &
$4.08^{+2.63}_{-1.79}$ & $2.54^{...}_{...}$ & 7.7 & 1.30 & 6.9\\
\noalign{\smallskip}
IC\,5152 & $2.35^{+2.27}_{-1.24}$ & $1.63^{+0.99}_{-1.24}$ & 5.3 & 1.17 &
$5.38^{+0.10}_{-0.64}$ & $5.38^{+0.10}_{-0.64}$ & 12.2 & 1.07 & 3.9\\
\noalign{\smallskip}
\noalign{\smallskip}
\hline
\end{tabular}
$$
\footnotesize{Notes: (1) the name of the galaxy; (2) the median of the
  quadratically-added amplitude of the non-circular motions (unconstrained
  case), averaged over the entire radial range, the errors indicate the lower
  and upper quartile; (3) same as (2), but averaged over the inner 1\,kpc
  only; (4) the percentage the non-circular motions contribute to the maximum
  rotation velocity; (5) the median of the \emph{reswri} absolute residual
  velocity field after the harmonic decomposition; (6), (7), (8), (9) the same
  as (2), (3), (4), (5), but for the constrained case; (10) maximum radius for
  the averaging of the quadratically-added amplitudes.}
\end{table*}

\begin{figure}
 \centering
\includegraphics[width=.49\textwidth,viewport=63 519 275 723, clip=]{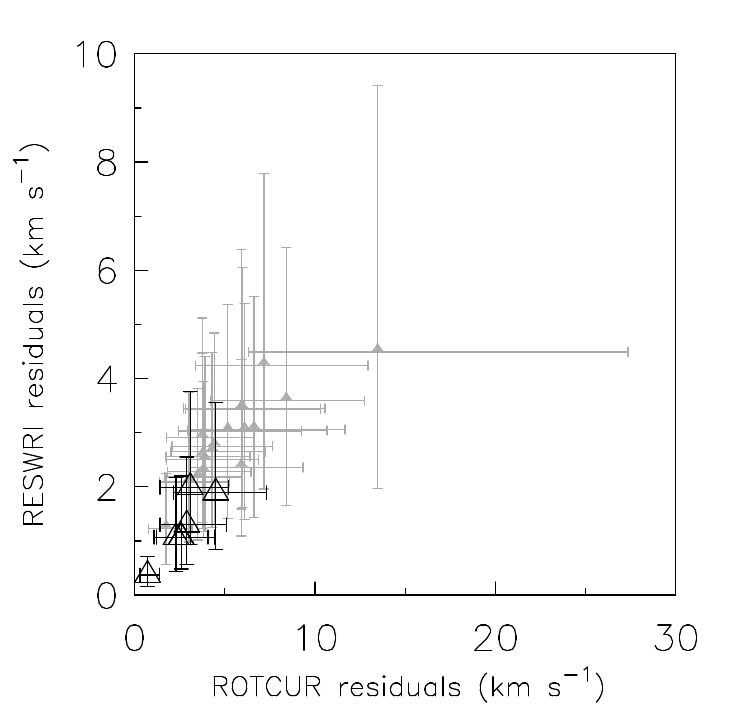}
\caption[]{Median of the absolute residual field from the harmonic
  decomposition (constrained case) \emph{vs.} the one from the rotation curve
  analysis. The light grey triangles represent the values derived by
  CT08. They are complemented by the results of our analysis (black open
  triangles). The error bars indicate the lower and upper quartile.}
\label{res}
\end{figure}

In order to check if our harmonic decomposition is able to quantify most of
the non-circular motions, we follow CT08 and make use of the residual velocity
fields. In Fig.~\ref{res} we compare the median of the absolute residual
velocity fields created with \emph{reswri} (constrained case) with the median
of the absolute residual velocity fields created with \emph{rotcur}. The
galaxies from our sample are presented by open black triangles. For a
comparison we also show the data points from CT08 (filled light grey
triangles). In agreement with CT08, the \emph{rotcur} residuals are larger
than the \emph{reswri} ones as \emph{rotcur} does not take non-circular
motions into account. In comparison to most of the galaxies studied by CT08,
our sample dwarf galaxies lie at the lower end of the distribution. The median
value of the \emph{reswri} residuals $<M_{\rm res}>$ is below 2\skms\ (see
Table~\ref{tabharm}), which implies that the harmonic decomposition up to
third order has captured most of the non-circular motions.

Altogether, we could show that while large non-circular motions might be found
in dwarf galaxies, they are the exception rather than the rule. The galaxies
in our sample have non-circular motions that are that small that they do not
significantly affect the rotation curves. This implies that non-circular
motions cannot artificially flatten the slope of the density profile and turn
a cusp into a core.
\section{Mass models}
The \HI\ rotation curves of our sample galaxies are only marginally influenced
by non-circular motions. Therefore, we use the ones derived with
  \emph{rotcur} (see Sect.~\ref{Sectvelorot} and Fig.~\ref{velomap_rot},
  middle panels) to perform a mass decomposition with the GIPSY task
\emph{rotmas}. This task allows to interactively fit the different components
(halo, gas, stars) to the observed rotation curve by minimising the $\chi^2$
of the parameter space. In this section, the dark matter halo models are
introduced and the derivation of the mass components is described.
\subsection{Dark matter halo models}
We use two models to describe the dark matter halo, which are the cuspy NFW
halo \citep[e.g.,][]{Navarro1996} and the cored pseudo-isothermal (ISO) halo
\citep[e.g.,][]{Binney1987}.

Numerical simulations show that the density of CDM haloes rises steeply
towards the halo centre \citep[e.g.,][]{Navarro1996}. The NFW mass-density
distribution is described as
\begin{equation}
\rho_{\rm NFW}(r)=\frac{\rho_i}{r/r_{\rm s} \cdot \left(1+r/r_{\rm s}\right)^2},
\end{equation}
where $\rho_{\rm i}$ is related to the density of the Universe at the time of
halo collapse, and $r_{\rm s}$ is the characteristic radius of the halo. This
leads to a rotation curve of
\begin{equation}
\vrot (r)=v_{200} \sqrt{\frac{\ln(1+cx)-(cx)/(1+cx)}{x[\ln(1+c)-c/(1+c)]}},
\end{equation}
with $x=r/r_{200}$. The concentration parameter $c=r_{200}/r_{\rm s}$ is
directly correlated to $v_{200}$, the circular velocity at $r_{200}$
\citep{deBlok2003}
\begin{equation}
\log c=1.191-0.064\,\log v_{200}-0.032\,\log v_{200}^2,
\label{deBlok}
\end{equation}
where $r_{200}$ is the radius at which the density contrast exceeds 200, i.e.,
roughly the virial radius \citep{Navarro1996}. Note that we do not
  correct the NFW fits for adiabatic compression. As \citet{Sellwood2005}
  show, the central halo concentration of dwarf galaxies increases only
  slightly from the primordial to the compressed halo, which means that it
  still fits the result from N-body simulations ($\alpha=1$).

As a second model the empirically derived ISO halo is used
\citep{Binney1987}. It describes a dark matter halo that has a core of roughly
constant density. The density profile is given by
\begin{equation}
\rho_{\rm ISO}(r)=\rho_0\left(1+\left(\frac{r}{r_{\rm c}}\right)^2\right)^{-1},
\end{equation}
with $\rho_0$ being the central density and $r_{\rm c}$ the core radius. The
rotation curve corresponding to this density profile can be expressed as
\begin{equation}
\vrot(r)=\sqrt{4 \pi G \rho_0 r_{\rm c}^2\left(1+\frac{r_{\rm c}}{r} \arctan \left(\frac{r}{r_{\rm c}}\right)\right)}.
\end{equation}
\subsection{Fitting process}
\label{Sectcases}
As dwarf galaxies are believed to be dark matter dominated at all radii
\citep{deBlok1997}, the velocity contribution from the baryons is often
neglected when decomposing the rotation curves of these galaxies. But although
dark matter is the dominant component in dwarf galaxies, the baryons, i.e.,
gas and stars, are still important, especially close to the centre. Including
all components, the observed rotation velocity is then given by
\begin{equation}
v_{\rm obs}^2=v_{\rm stars}^2+v_{\rm gas}^2+v_{\rm halo}^2.
\end{equation}
We model the rotation curves in various ways, partly including the
baryons. The different approaches are described in the following.
\subsubsection{Minimum-disc case}
This is the simplest case ignoring all baryonic contributions. Thus, only the
dark matter haloes are fitted to the observed rotation curve.
\subsubsection{Minimum-disc + gas case}
In this case, the contribution of the gas is taken into account as well. Next
to \HI, which dominates the gas component, He and metals are included by
scaling the \HI\ column density by a factor of 1.5. As the ratio between the
molecular gas and \HI\ is much lower in dwarf galaxies than in luminous spirals
\citep[][]{Taylor1998, Leroy2005}, we do not correct for this gas
component. In order to create a surface density profile for the \HI, the GIPSY
task \emph{ellint} is taken, using the \HI\ intensity maps
(Figs.~\ref{N2366harm} to \ref{IC5152harm}, lower right panels) and the
parameters from the best tilted-ring model (see Table~\ref{Kinpar}) as an
input. The output of \emph{ellint} is given in terms of mean flux and has
therefore to be converted to physical units. We then use the GIPSY task
\emph{rotmod} in order to determine the rotation of the gas under the
assumption of an infinitesimally thin disc.
\subsubsection{Maximum-disc case}
\label{Maxdisc}
Here, the contribution of the gas, the stars and the dark matter halo are
simultaneously fitted to the observed rotation curve. For all galaxies, the
surface density can be described by a S\'{e}rsic model \citep{Sersic1968}
\begin{equation}
 \mu=\mu_0+1.086\,\left (\frac{r}{h}\right)^n,
\end{equation}
with $\mu_0$ being the central surface brightness, \emph{h} the disc scale
length, and \emph{n} the shape parameter (where $n=1$ gives an exponential
profile). The parameters for each galaxy are listed in
Table~\ref{expdiscparams}. They are converted from $\rm mag/arcsec^2$ to units
of $\rm L_{\odot}/pc^2$. 
Again, we use \emph{rotmod} to determine the contribution of the stars under
the assumption of an infinitesimally thin disc. The derivation of the
contribution of the stellar disc is the most critical part: first of all, the
surface photometry depends on different factors which are difficult to
estimate like, e.g., the extinction. Secondly, in order to derive the stellar
rotation curve, the mass to light ratio \emph{M/L} has to be known. Note that
we make the assumption of a stellar \emph{M/L} of 1 in the case
of NGC\,2366 and of 2 in the case of ESO\,215-G?009, which are in good
agreement with adopting a Kroupa or a Kennicutt initial mass function
\citep[][and references therein]{Portinari2004}. For the other sample
galaxies, calculations by assuming a Kroupa or Kennicutt initial mass function
reveal stellar \emph{M/L} ratios that result in a stellar rotation curve which
lies above the observed one. Therefore, we choose smaller stellar \emph{M/L}
ratios, but always greater than 0.2, which is still a physical value
\citep[see also][]{Spano2008}. In case of NGC\,5408, we do not fit a stellar
component as the \emph{M/L} value has to be far below 0.1 in order to fit the
observed rotation curve.
\begin{table*}
\centering 
\caption[Parameters for the exponential disc fitting.]{Parameters for the exponential disc fitting.}
\label{expdiscparams}
$$
\begin{tabular}{lccccccc}
\hline
  \hline
\noalign{\smallskip}
Parameter [Unit] & NGC\,2366 & ESO\,059-G001 & ESO\,215-G?009 & NGC\,4861 &
  NGC\,5408 & IC\,5152\\
& (1) & (2) & (3) & (4) & (5) & (2)\\
\hline
\noalign{\smallskip}
Band & \emph{V} & \emph{H} & \emph{R} & \emph{R} & \emph{J} & \emph{H}\\
$\mu_0\,[\rm mag/arcsec^2]$ & 22.82 & 20.23 & 23.16 & 22.95 & 18.29 & 17.65\\
$h$ [kpc] & 1.59 & 0.81 & 2.16 & 0.88 & 0.33 & 0.36\\
$n$ & 1 & 1.33 & 1 & 1 & 1 & 0.96\\
\noalign{\smallskip}
\hline
\noalign{\smallskip}
$M/L$ & 1 & 0.8 & 2 & 0.3 & $<0.1$ & 0.2\\
\noalign{\smallskip}
\hline
\end{tabular}
$$
\flushleft {\footnotesize References: photometry by (1) \citet{Hunter2001}, (2) \citet{Kirby2008}, (3) \citet{Warren2006}, (4) \citet{GildePaz2005}, (5) \citet{Noeske2003}.}
\end{table*}
\section{Results and discussion}
As described in, e.g., vE09a, the outer parts of the rotation curves show
large uncertainties due to the sparsely filled tilted rings. Therefore, we
decompose the rotation curves twice for each of the different cases (see
Sect.~\ref{Sectcases}) and for each dark matter density profile: first, using
the entire rotation curve and secondly, using a truncated rotation curve
restricted to the inner few kpcs only. The results for all sample galaxies
are shown in Figs.~\ref{N2366dmh} to \ref{IC5152dmh}. From top to bottom the
minimum-disc, the minimum-disc\,+\,gas and the maximum-disc cases are
fitted. Solid triangles represent the observed rotation curve, long-dashed
lines the dark matter halo component, dotted lines the gas component, and
dashed-dotted lines the contribution of the stars. The resulting model fit is
indicated by a solid line.

Tables~\ref{Massmodelsnfw} and \ref{Massmodelsiso} list the results of the
best fits for both halo profiles including the reduced $\chi^2$ values. We
assume $v_{200}$ to be the rotation speed, which we estimate from the flat
parts of the observed rotation curves. Following Eq.~\ref{deBlok}, we then
derive the concentration parameter c, which is between 9 and 10 for all
sample galaxies, i.e., consistent with CDM simulations. Both parameters are
kept fixed so that $r_{200}$ is the only free parameter. In the case of the
ISO halo both parameters, the core radius $r_{\rm c}$ and the core density
$\rho_{\rm 0}$, are allowed to vary.

In the following subsections we present the results for each galaxy separately.
\begin{table*}
\caption{The basic parameters of the mass decomposition - the NFW model.}
\label{Massmodelsnfw}
$$
\begin{tabular}{lcccccccccc}
  \hline
  \hline
  \noalign{\smallskip}
  & \multicolumn{5}{c}{NFW full} & \multicolumn{5}{c}{NFW small}\\
  Parameter & $v_{200}$ & c & $r_{200}$ & $\Delta r_{200}$ & $\chi^2_{\rm red}$ & $v_{200}$ & c & $  r_{200}$ & $\Delta r_{200}$ & $\chi^2_{\rm red}$\\
  Unit & [\kms] & & [kpc] & [kpc] & & [\kms] & & [kpc] & [kpc] & \\
  & (1) & (2) & (3) & (4) & (5) & (6) & (7) & (8) & (9) & (10)\\
  \hline
  \noalign{\smallskip}
  \multicolumn{11}{c}{NGC\,2366}\\
  \hline
  \noalign{\smallskip}
  Min.-disc & 50 & 9.4 & 117.54 & 7.26 & 47.47 & 50 & 9.4 & 92.81 & 4.52 & 19.88\\
  Min.-disc\,+\,gas & 50 & 9.4 & 206.12 & 16.44 & 51.76 & 50 & 9.4 & 136.75 &
  5.06 & 8.19\\
  Max.-disc & 50 & 9.4 & 386.29 & 37.63 & 36.27 & 50 & 9.4 & 229.04 & 11.39 & 7.81\\
  \hline
  \noalign{\smallskip}
  \multicolumn{11}{c}{ESO\,059-G001}\\
  \hline
  \noalign{\smallskip}
  Min.-disc & 69 & 9.2 & 89.22 & 7.84 & 22.81 & ... & ... & ... & ... & ...\\
  Min.-disc\,+\,gas & 69 & 9.2 & 96.00 & 8.04 & 19.67 & ... & ... & ... & ... & ...\\
  Max.-disc & 69 & 9.2 & 124.40 & 21.27 & 63.70 & ... & ... & ... & ... & ...\\
  \hline
  \noalign{\smallskip}
  \multicolumn{11}{c}{ESO\,215-G?009}\\
  \hline
  \noalign{\smallskip}
  Min.-disc & 77 & 9.0 & 134.57 & 5.13 & 19.07 & 77 & 9.0 & 128.08 & 3.63 & 9.14\\
  Min.-disc\,+\,gas & 77 & 9.0 & 157.35 & 7.85 & 30.18 & 77 & 9.0 & 145.41 &
  5.00 & 12.48\\
  Max.-disc & 77 & 9.0 & 309.86 & 17.19 & 19.95 & 77 & 9.0 & 283.81 & 11.67 & 9.13\\
  \hline
  \noalign{\smallskip}
  \multicolumn{11}{c}{NGC\,4861}\\
  \hline
  \noalign{\smallskip}
  Min.-disc & 46 & 9.5 & 125.79 & 10.55 & 17.42 & 46 & 9.5 & 132.55 & 18.92 & 29.44\\
  Min.-disc\,+\,gas & 46 & 9.5 & 181.48 & 13.27 & 10.93 & 46 & 9.5 & 189.54 &
  24.43 & 16.57\\
  Max.-disc & 46 & 9.5 & 219.91 & 26.36 & 24.86 & 46 & 9.5 & 267.28 & 65.64 & 37.31\\
  \hline
  \noalign{\smallskip}
  \multicolumn{11}{c}{NGC\,5408}\\
  \hline
  \noalign{\smallskip}
  Min.-disc & 53 & 9.7 & 182.70 & 24.22 & 47.97 & 53 & 9.7 & 187.81 & 34.65 & 63.22\\
  Min.-disc\,+\,gas & 53 & 9.7 & 233.18 & 30.95 & 38.76 & 53 & 9.7 & 239.76 &
  46.08 & 51.79\\
  \hline
  \noalign{\smallskip}
  \multicolumn{11}{c}{IC\,5152}\\
  \hline
  \noalign{\smallskip}
  Min.-disc & 44 & 10.0 & 34.26 & 1.80 & 4.92 & 44 & 10.0 & 31.03 & 1.90 & 4.51\\
  Min.-disc\,+\,gas & 44 & 10.0 & 43.47 & 2.72 & 7.09 & 44 & 10.0 & 37.09 &
  2.52 & 5.12\\
  Max.-disc & 44 & 10.0 & 79.97 & 5.57 & 6.93 & 44 & 10.0 & 90.67 & 13.45 & 10.58\\
  \hline
  \noalign{\smallskip}
  \end{tabular}
$$
\footnotesize{Notes: (1) and (6): the rotation velocity at $r_{200}$, estimated from the flat part of the rotation curve; (2) and (7):
  the concentration parameter derived following Eq.~\ref{deBlok}; (3) and (8):
  the fitted virial radius; (4) and (9): uncertainties of $r_{200}$; (5) and
  (10): reduced $\chi^2$ values.}
\end{table*}
\begin{table*}
\caption{The basic parameters of the mass decomposition - the
  pseudo-isothermal halo model.}
\label{Massmodelsiso}
$$
\begin{tabular}{lcccccccccc}
  \hline
  \hline
  \noalign{\smallskip}
& \multicolumn{5}{c}{ISO full} & \multicolumn{5}{c}{ISO small}\\
Parameter & $\rho_0$ & $\Delta \rho_0$ & $r_{\rm c}$ & $\Delta r_{\rm c}$ & $\chi^2_{\rm red}$ & $\rho_0$ & $\Delta \rho_0$ & $r_{\rm c}$ & $\Delta r_{\rm c}$ & $\chi^2_{\rm red}$\\
Unit & $\rm [10^{-3} M_\odot\,pc^{-3}]$ & $\rm [10^{-3} M_\odot\,pc^{-3}]$ &
[kpc] & [kpc] & & $\rm [10^{-3}
  M_\odot\,pc^{-3}]$ & $\rm [10^{-3} M_\odot\,pc^{-3}]$ & [kpc] & [kpc]\\
  & (1) & (2) & (3) & (4) & (5) & (6) & (7) & (8) & (9) & (10)\\
  \hline
  \noalign{\smallskip}
  \multicolumn{11}{c}{NGC\,2366}\\
  \hline
  \noalign{\smallskip}
  Min.-disc & 56.66 & 10.42 & 0.94 & 0.11 & 27.82 & 32.95 & 1.33 & 1.58 &
  0.05 & 1.79\\
  Min.-disc\,+\,gas & 53.12 & 14.46 & 0.77 & 0.13 & 30.27 & 27.54 & 1.02 &
  1.44 & 0.04 & 0.88\\
  Max.-disc & 25.28 & 9.82 & 0.92 & 0.22 & 24.99 & 11.27 & 0.72 & 2.23 & 0.14 & 1.24\\
  \hline
  \noalign{\smallskip}
  \multicolumn{11}{c}{ESO\,059-G001}\\
  \hline
  \noalign{\smallskip}
  Min.-disc & 67.08 & 3.18 & 1.48 & 0.06 & 0.84 & ... & ... & ... & ... & ...\\
  Min.-disc\,+\,gas & 62.98 & 3.58 & 1.49 & 0.07 & 1.10 & ... & ... & ... & ... & ...\\
  Max.-disc & 25.70 & 5.37 & 3.03 & 0.76 & 12.52 & ... & ... & ... & ... & ...\\
  \hline
  \noalign{\smallskip}
  \multicolumn{11}{c}{ESO\,215-G?009}\\
  \hline
  \noalign{\smallskip}
  Min.-disc & 77.74 & 10.73 & 1.27 & 0.11 & 18.42 & 71.67 & 9.27 & 1.35 & 0.12
  & 15.67\\
  Min.-disc\,+\,gas & 98.86 & 15.22 & 1.01 & 0.10 & 16.47 & 89.34 & 11.62 &
  1.10 & 0.09 & 11.77\\
  Max.-disc & 34.27 & 8.49 & 1.47 & 0.24 & 19.18 & 28.47 & 6.10 & 1.71 & 0.26 & 14.44\\
  \hline
  \noalign{\smallskip}
  \multicolumn{11}{c}{NGC\,4861}\\
  \hline
  \noalign{\smallskip}
  Min.-disc & 13.98 & 1.58 & 2.22 & 0.18 & 4.29 & 11.06 & 0.70 & 3.04 & 0.22 & 1.00\\
  Min.-disc\,+\,gas & 9.67 & 1.29 & 2.45 & 0.24 & 3.91 & 8.75 & 0.98 & 2.90 & 0.35 & 1.72\\
  Max.-disc & 4.85 & 1.13 & 3.93 & 0.81 & 10.40 & 3.95 & 1.48 & 7.10 & 7.19 & 14.21\\
  \hline
  \noalign{\smallskip}
  \multicolumn{11}{c}{NGC\,5408}\\
  \hline
  \noalign{\smallskip}
  Min.-disc & 9.22 & 2.11 & 3.36 & 0.68 & 20.68 & 6.69 & 0.87 & 6.37 & 1.66 &
  6.85\\
  Min.-disc\,+\,gas & 6.98 & 1.76 & 3.70 & 0.87 & 18.13 & 4.94 & 0.78 & 8.64 &
  4.05 & 7.24\\
  \hline
  \noalign{\smallskip}
  \multicolumn{11}{c}{IC\,5152}\\
  \hline
  \noalign{\smallskip}
  Min.-disc & 378.79 & 40.76 & 0.37 & 0.02 & 1.94 & 389.45 & 40.23 & 0.36 &
  0.02 & 1.24\\
  Min.-disc\,+\,gas & 369.69 & 48.16 & 0.34 & 0.03 & 2.23 & 398.06 & 49.18 &
  0.33 & 0.03 & 1.36\\
  Max.-disc & 42.11 & 6.70 & 1.18 & 0.14 & 2.87 & 47.24 & 15.81 & 1.08 & 0.36 & 4.44\\
  \hline
  \noalign{\smallskip}
\end{tabular}
$$
\footnotesize{Notes: (1) and (6): fitted core density $\rho_0$ of the
  pseudo-isothermal halo model; (2) and (7): uncertainties of $\rho_0$; (3)
  and (8): fitted core $r_{\rm c}$ radius of the pseudo-isothermal halo model;
  (4) and (9): uncertainties of $r_{\rm c}$; (5) and (10): reduced $\chi^2$
  values.}
\end{table*}
\subsection{NGC\,2366}
A mass decomposition of NGC\,2366 has already been done by Oh08. They derived
the rotation curve from a so-called ``bulk'' velocity field, which only
contains the circularly rotating components (Oh08). Our analysis is done on
the Hermite velocity field that still contains non-circular motions. However,
as shown in Sect.~\ref{SectN2366harm}, the non-circular motions in NGC\,2366
(as well as in the other sample galaxies) do not significantly affect the
rotation curves.

\begin{figure*}
\centering
\includegraphics[width=.9\textwidth,viewport= 58 353 520 653,clip=]{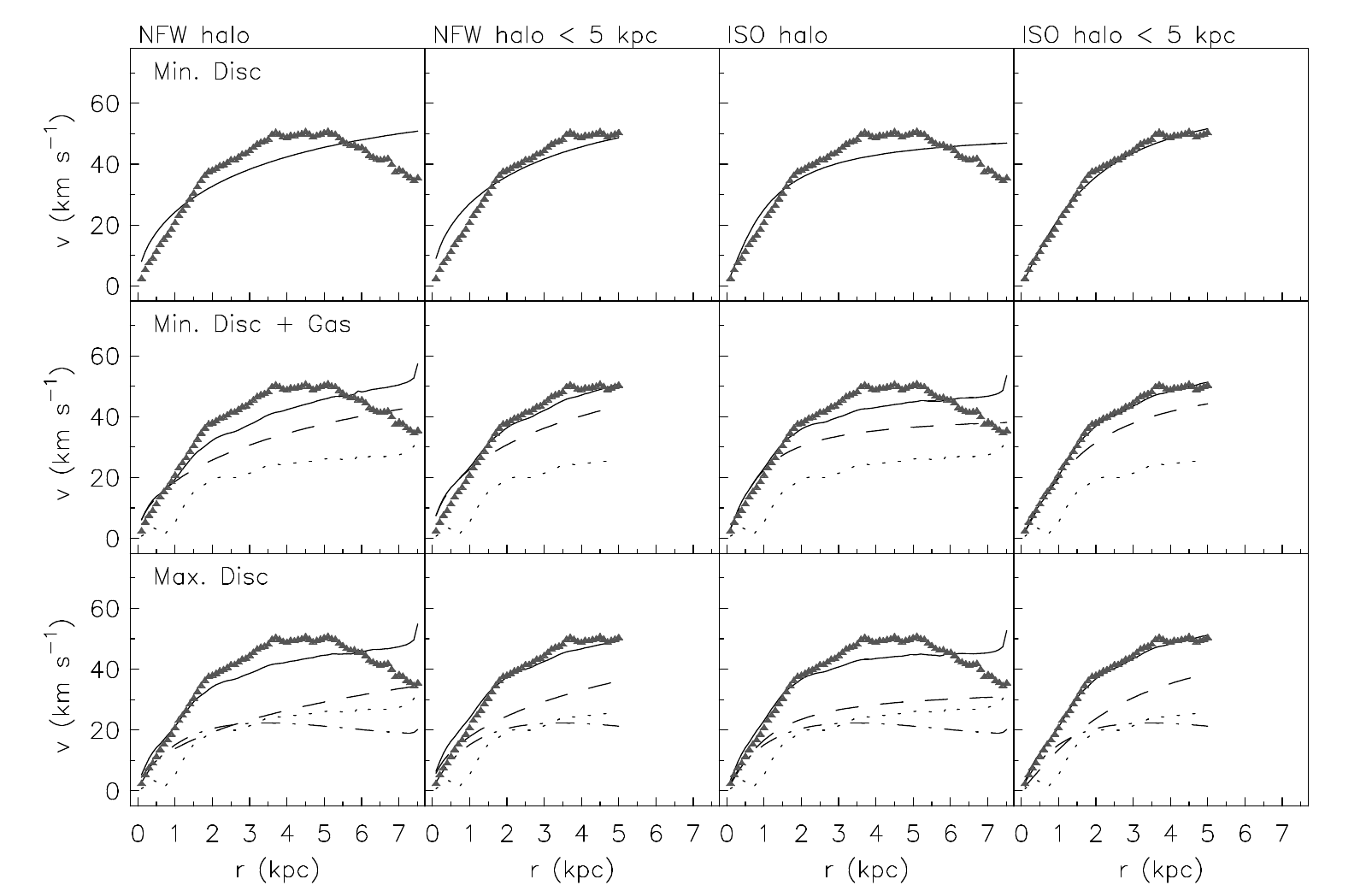}
\caption[Mass decomposition of NGC\,2366.]{Mass decomposition of
  NGC\,2366. The rotation curves for an NFW halo and a pseudo-isothermal halo
  are modelled for the full rotation curve, and for the inner 5\,kpc
  only. From top to bottom the minimum-disc, the minimum-disc\,+\,gas, and the
  maximum-disc cases are fitted. Solid triangles represent the observed
  rotation curve, long-dashed lines the dark matter halo component, dotted
  lines the gas component, dashed-dotted lines the contribution of
  the stars, and solid lines the resulting model fit.}
\label{N2366dmh}
\end{figure*}

Figure~\ref{N2366dmh} shows the resulting model fits. A comparison of the
reduced $\chi^2$ values (see Tables~\ref{Massmodelsnfw} and
\ref{Massmodelsiso}) reveals that the ISO halo gives much better results than
the NFW halo independent of the radius of the rotation curve. In the outer
parts, the observed rotation curve is affected by uncertainties caused by the
sparsely filled tilted rings and by asymmetries in the \HI\ velocity field
(cf. vE09a). Therefore, it strongly declines from a radius of 5\,kpc on so
that we also model the curve within the inner 5\,kpc only. This gives improved
fits for both haloes: $\chi^2_{\rm red}$ is lower by a factor of four when
using the NFW halo and even by a factor of 20 when using the ISO
halo. Additionally, $\chi^2_{\rm red}$ decreases by including gas and stars.

The reduced \HI\ rotation curve fitted with an ISO halo gives the lowest
values for $\chi^2_{\rm red}$. The resulting model fit agrees very well with
the observed curve. Moreover, the values of the core
densities and radii from the ISO models are comparable to the results of Oh08.
\subsection{ESO\,059-G001}
\begin{figure*}
\centering
\includegraphics[width=.5\textwidth,viewport= 58 343 320 653,clip=]{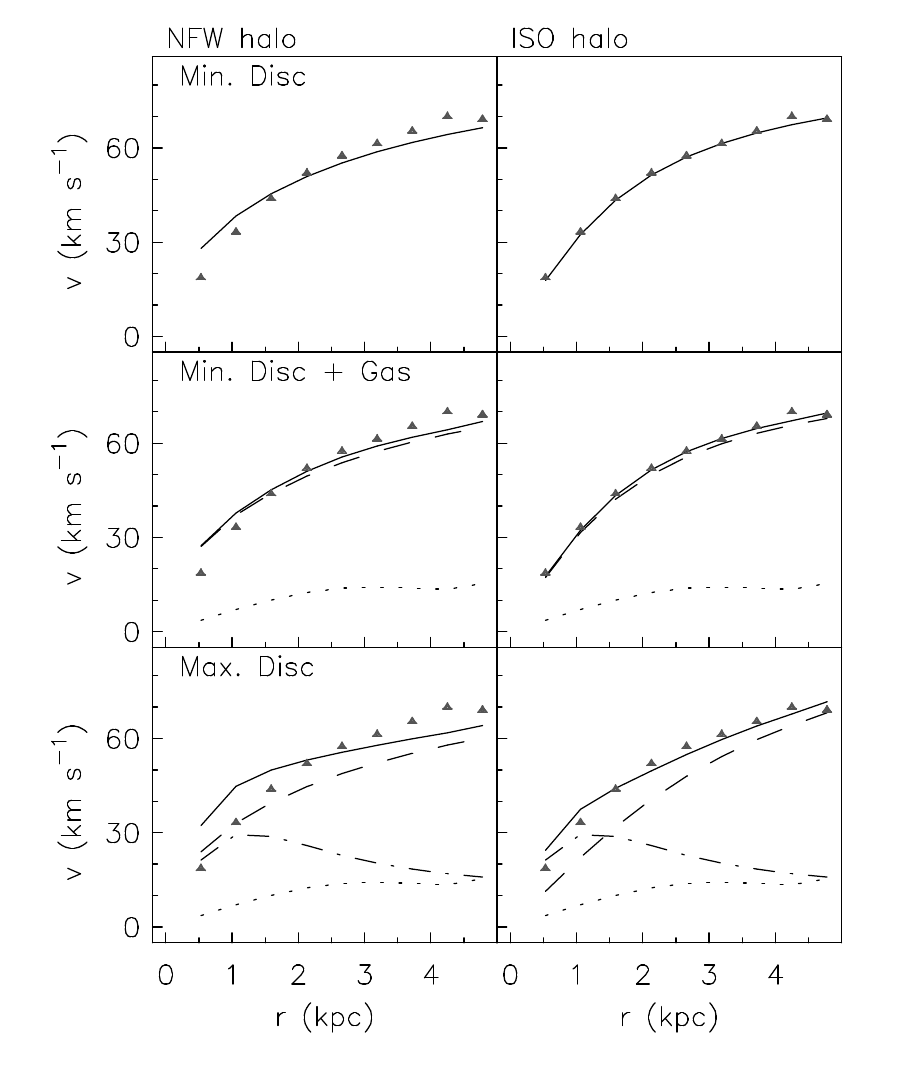}
\caption{Mass decomposition of ESO\,059-G001. The same as in
  Fig.~\ref{N2366dmh}. No reduced rotation curve is fitted.}
\label{ESO059-G001dmh}
\end{figure*}

As the rotation velocities of this galaxy have small errors (also in the outer
parts), we only perform a mass decomposition for the entire rotation
curve. Figure~\ref{ESO059-G001dmh} shows the resulting curves. Again, the ISO
halo gives much better results than the NFW halo. Especially the models for
the minimum-disc and minimum-disc\,+\,gas cases using the truncated rotation
curve result in very low $\chi^2_{\rm red}$ values (see
Tables~\ref{Massmodelsnfw} and \ref{Massmodelsiso}).
\subsection{ESO\,215-G?009}
\label{SectESO215-G?009massdecomp}
\begin{figure*}
\centering
\includegraphics[width=.9\textwidth,viewport= 58 343 520 653,clip=]{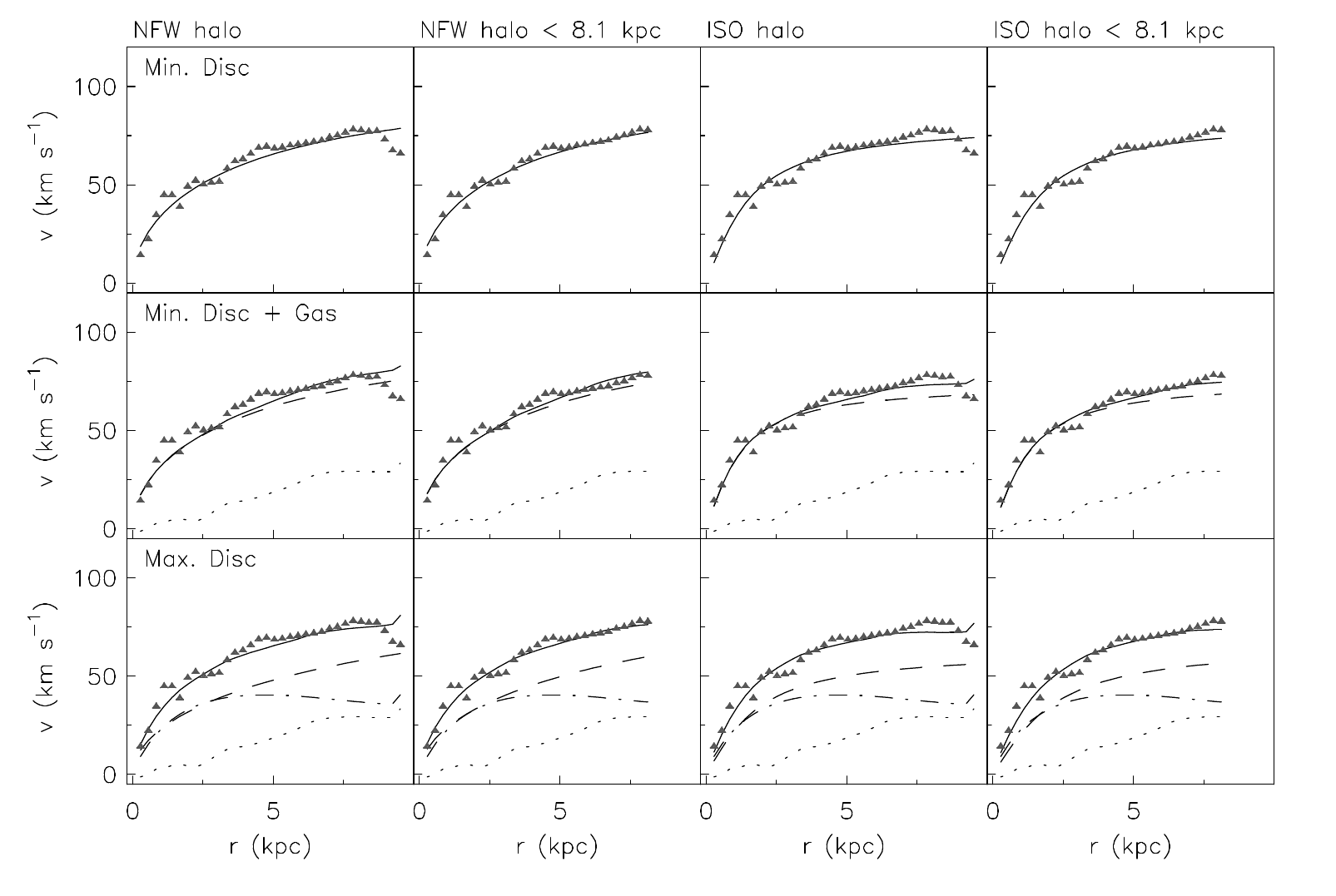}
\caption{Mass decomposition of ESO\,215-G?009. The same as in
  Fig.~\ref{N2366dmh}. A reduced rotation curve is fitted up to a radius of
  8.1\,kpc.}
\label{ESO215-G?009dmh}
\end{figure*}

ESO\,215-G?009 is modelled over the entire radial range and over the inner
8.1\,kpc only (see Fig.~\ref{ESO215-G?009dmh}). In both cases, the NFW halo
fits sometimes even give lower $\chi^2_{\rm red}$ values than the ISO halo
fits. This can probably be explained by the slightly scattered rotation
velocities between 1 and 3\,kpc.
\subsection{NGC\,4861}
\begin{figure*}
\centering
\includegraphics[width=.9\textwidth,viewport= 58 383 520 693,clip=]{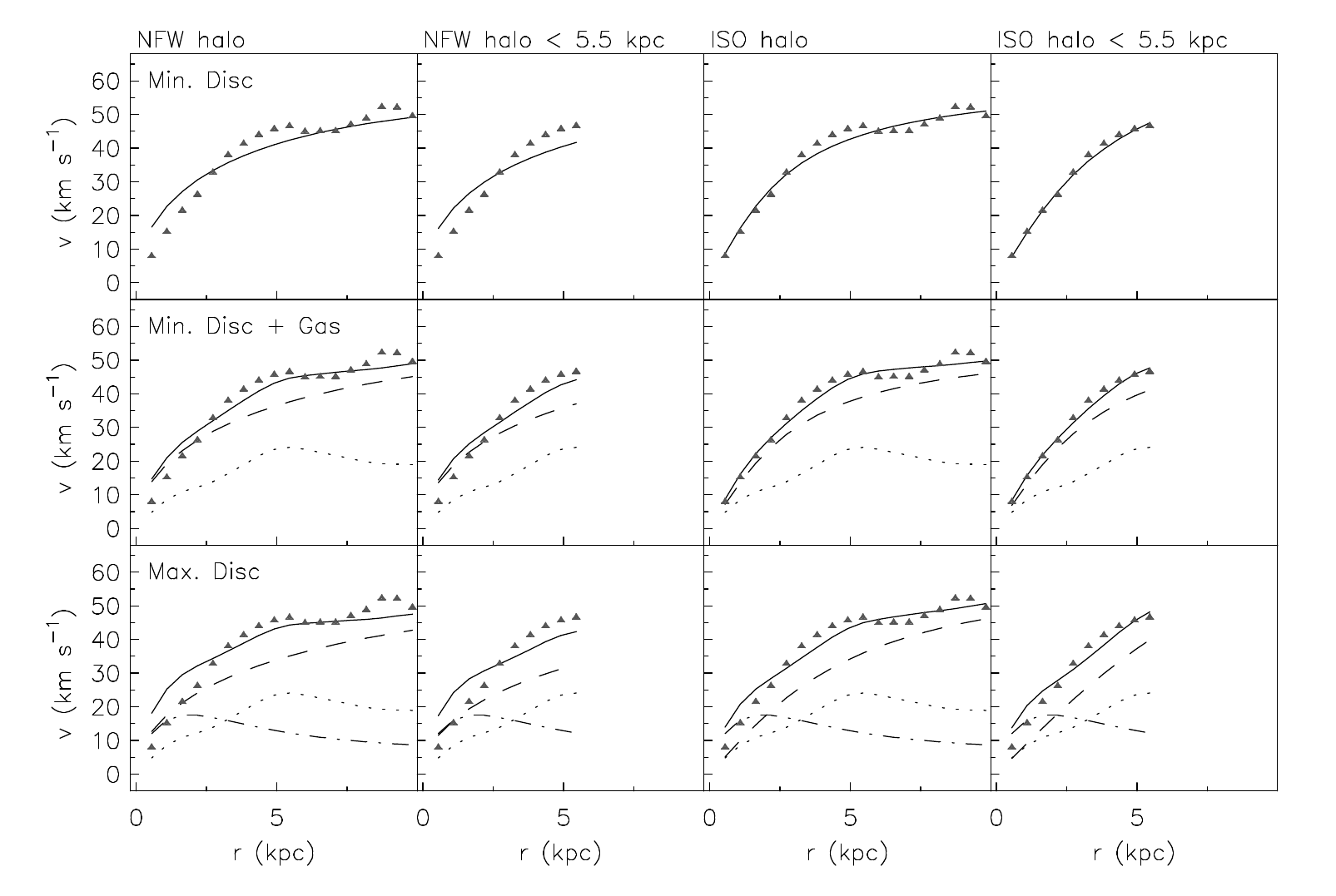}
\caption[Mass decomposition of NGC\,4861.]{Mass decomposition of
  NGC\,4861. The same as in Fig.~\ref{N2366dmh}. A reduced rotation curve is
  fitted up to a radius of 5.5\,kpc.}
\label{N4861dmh}
\end{figure*}

Figure~\ref{N4861dmh} shows the results for NGC\,4861. The reduced $\chi^2$
values of the maximum-disc case reveal that the stars are not well represented
by our estimate as the $\chi^2_{\rm red}$ values increase, independent of the
model, after adding the stellar component (see Tables~\ref{Massmodelsnfw} and
\ref{Massmodelsiso}). Similar to the
other galaxies in our sample the ISO matches the observed rotation curve
better than the NFW halo.
\subsection{NGC\,5408}
The \HI\ rotation curve of NGC\,5408 rises very slowly in the inner
2\,kpc. Using the photometric data from \citet{Noeske2003} (see
Table~\ref{expdiscparams}), it is impossible to derive a stellar rotation
curve that lies below the total rotation curve, even when assuming a stellar
\emph{M/L} ratio of 0.1, which is the lowest value before becoming unphysical
\citep{Spano2008}. We either underestimate the rotation in the central part of
the galaxy due to the low spatial resolution or the photometric data are not
accurate enough. We therefore decide against modelling the stellar component.

\begin{figure*}
\centering
\includegraphics[width=.9\textwidth,viewport= 58 436 520 653,clip=]{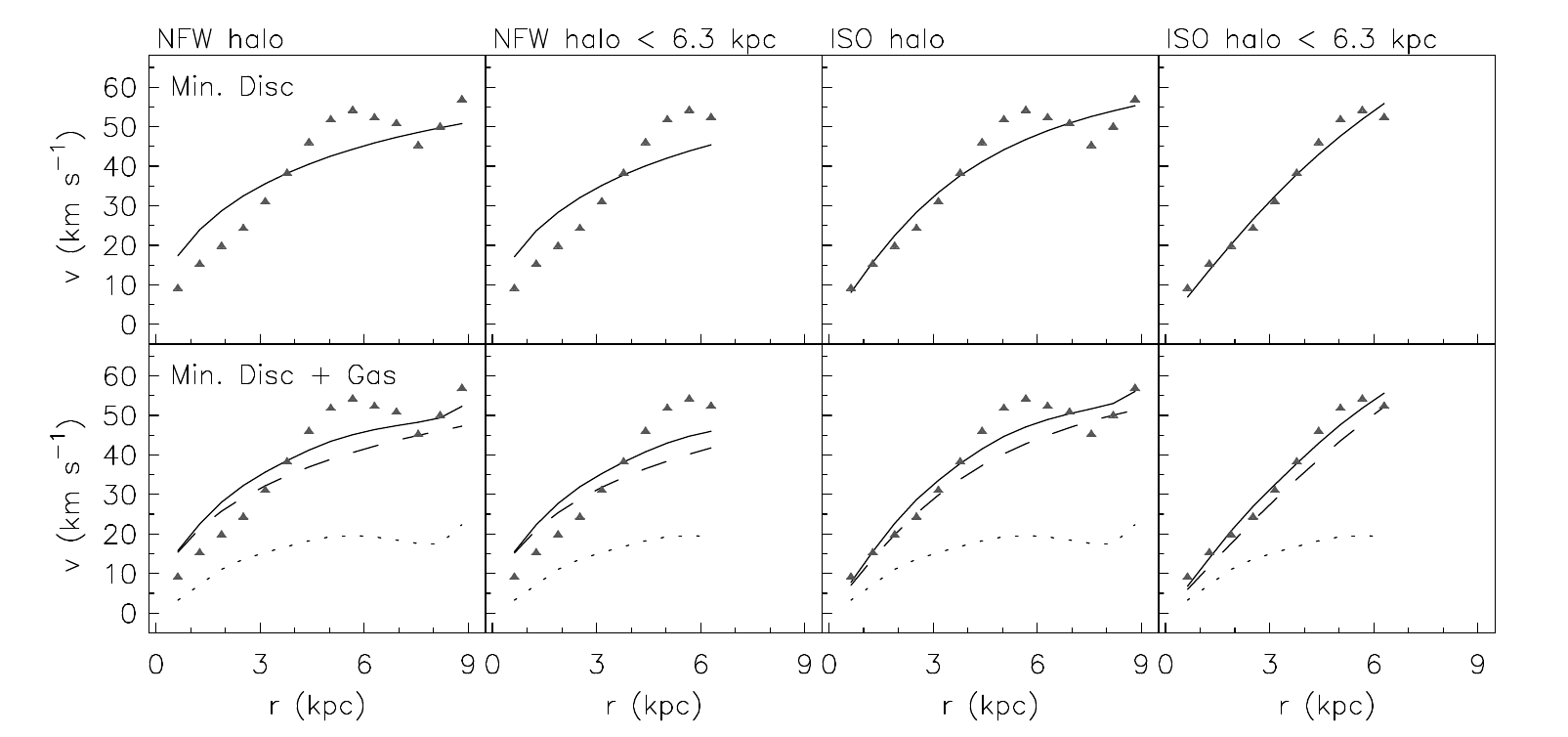}
\caption[Mass decomposition of NGC\,5408.]{Mass decomposition of
  NGC\,5408. The stellar contribution cannot be fitted. A reduced
  rotation curve is fitted up to a radius of 6.3\,kpc. Otherwise the same as in
  Fig.~\ref{N2366dmh}.}
\label{N5408dmh}
\end{figure*}

Figure~\ref{N5408dmh} shows the resulting fits. Again, the $\chi^2_{\rm red}$
values are very high for the NFW halo and much smaller for the ISO halo. For
this galaxy, the reduced rotation curve did not improve the quality of the
models, which is probably due to the fact that the repeated rise of the
rotation curve beyond a radius of 7\,kpc matches the first rise of the curve.
\subsection{IC\,5152}
\begin{figure*}
\centering
\includegraphics[width=.9\textwidth,viewport= 58 343 520 653,clip=]{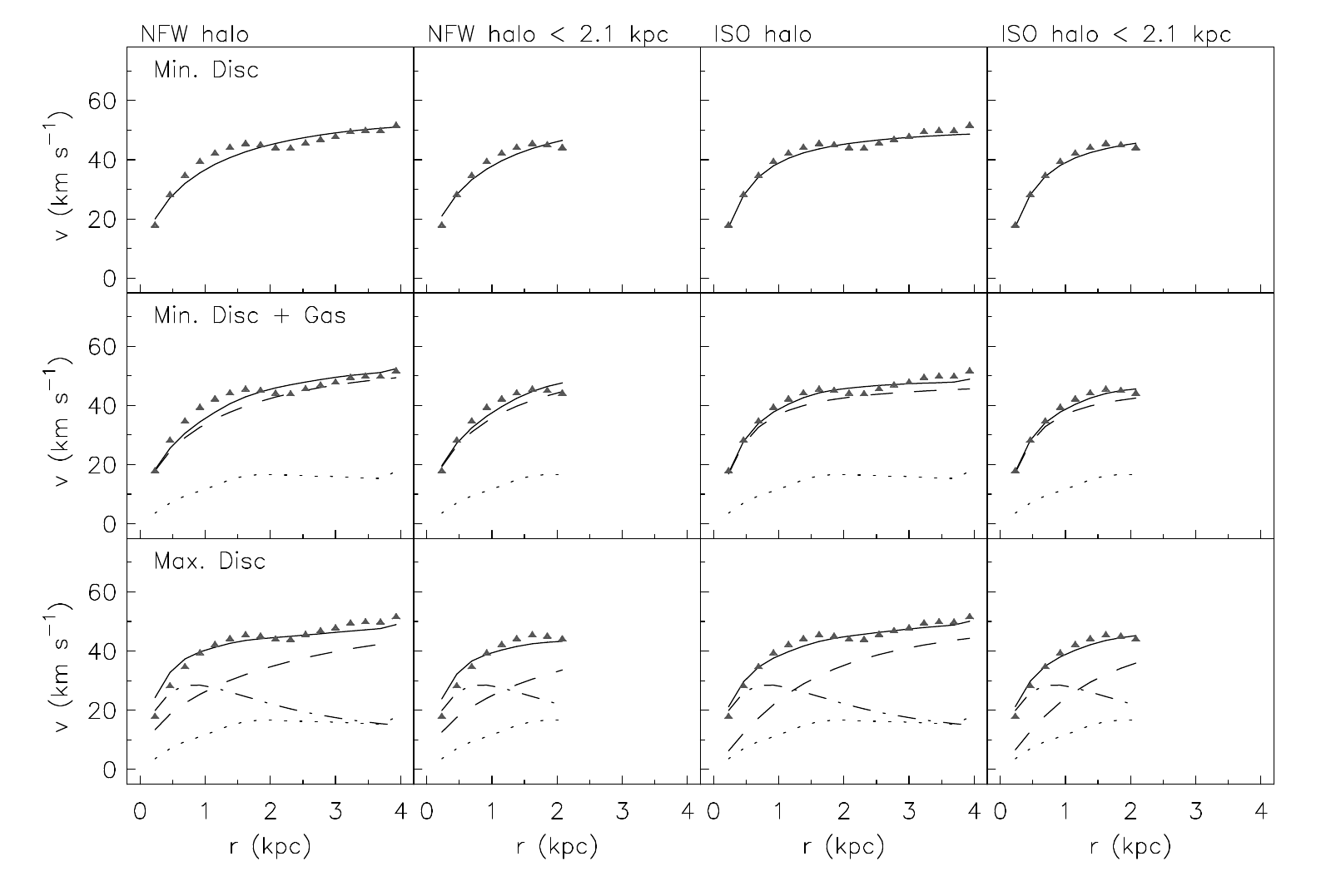}
\caption{Mass decomposition of IC\,5152. The same as in Fig.~\ref{N2366dmh}. A
  reduced rotation curve is fitted up to a radius of 2.1\,kpc.}
\label{IC5152dmh}
\end{figure*}

The results for IC\,5152 are presented in Fig.~\ref{IC5152dmh}. We fit NFW and
ISO haloes to the entire radial range and to the inner 2.1\,kpc only. Although
both dark matter halo models provide adequate fits (the $\chi^2_{\rm red}$
values for the NFW model are the lowest of our sample), the ISO model still
gives better values than the NFW halo.
\subsection{Discussion}
In contrast to recent work by dB08 and Oh08, we fitted the NFW profiles
  with fixed values of $v_{200}$ and c (see the beginning of this
  section). With this approach we assumed reasonable values for $v_{200}$ and
  c that follow cosmological predictions, and also got reasonable results for
  $r_{200}$. Nevertheless, we could show that in general, the observed rotation
  curves can better be fitted by an ISO halo than by an NFW halo, although at
  least in the case of ESO\,215-G?009 the NFW fits are equally good or even
  better than the ISO fits. The values of the core radius $r_{\rm c}$ and the
  core density $\rho_0$ are plausible and in good agreement with results for
  LSB galaxies \citep[][]{deBlok2001a,Kuzio2008} and for dwarf galaxies
\citep{Spano2008}. They also follow the recent finding of a constant
  central surface density $\mu_{\rm 0D}=r_{\rm c}\,\rho_0$ \citep[][and
    references therein]{Donato2009}.

\begin{figure*}
\centering
\includegraphics[width=.8\textwidth,viewport= 58 213 470 723,clip=]{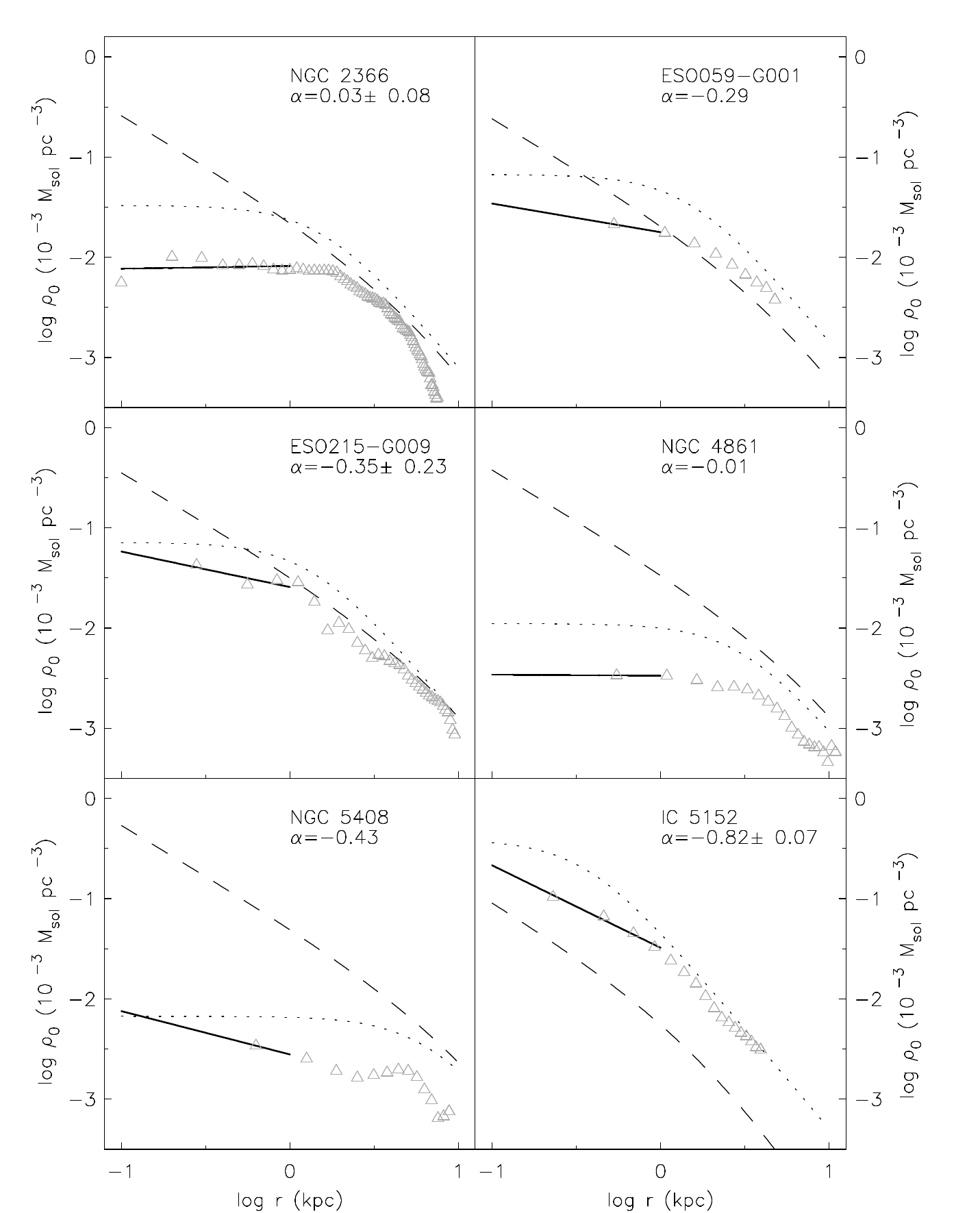}
\caption{The density profiles of all sample galaxies derived from the observed
  rotation curves (open grey triangles). Their inner slopes $\alpha$ are
  measured by applying a least square fit to all data points within the
  innermost kpc (bold black lines). The fitted values of $\alpha$ and the
  uncertainties are placed into the upper right corner of each panel. Note
  that the rotation curves of ESO\,059-G001, NGC\,4861, and NGC\,5408 only
  contain two points in the inner 1\,kpc. Therefore, no uncertainties can be
  given. The long-dashed and dotted lines show the NFW and the ISO profiles
  respectively using the parameters of the minimum-disc case.}
\label{density}
\end{figure*}

As mentioned in the introduction, CDM simulations predict density profiles
with a steep inner slope of $\alpha=-1$ \citep{Navarro1996} or even
$\alpha=-1.5$ \citep{Moore1998}. \citet{deBlok2001a} showed for their sample of
LSB galaxies that the observed distribution of $\alpha$ peaks at
$-0.2\pm0.2$. In order to compare these results with ours, we first measure
the slopes in the inner kpc of our data by plotting the logarithm of the
density \emph{vs.} the logarithm of the radius (see Fig.~\ref{density}). For
all galaxies except for IC\,5152, $\alpha$ lies between $-0.43$ and 0.03 (see
also Table~\ref{tabalpha}), which is within the errors of the $\alpha$
determined by \citet{deBlok2001a}. ESO\,215-G?009, which partly showed higher
$\chi^2_{\rm red}$ values for the ISO halo model than for the NFW halo model
(see Sect.~\ref{SectESO215-G?009massdecomp}), has an $\alpha$ of
$-0.35$. This means that its innermost slope can better be described by an
ISO halo.

In Fig.~\ref{density} we also fitted the ISO and NFW profiles of the minimum-disc case (dotted and long-dashed lines respectively). They do not always
agree with the observed density profiles, which is to be expected as
$\chi^2_{\rm red}$ is often quite high indicating a bad fit.

\begin{figure}
\centering
\includegraphics[width=.49\textwidth,viewport= 63 523 270 713,clip=]{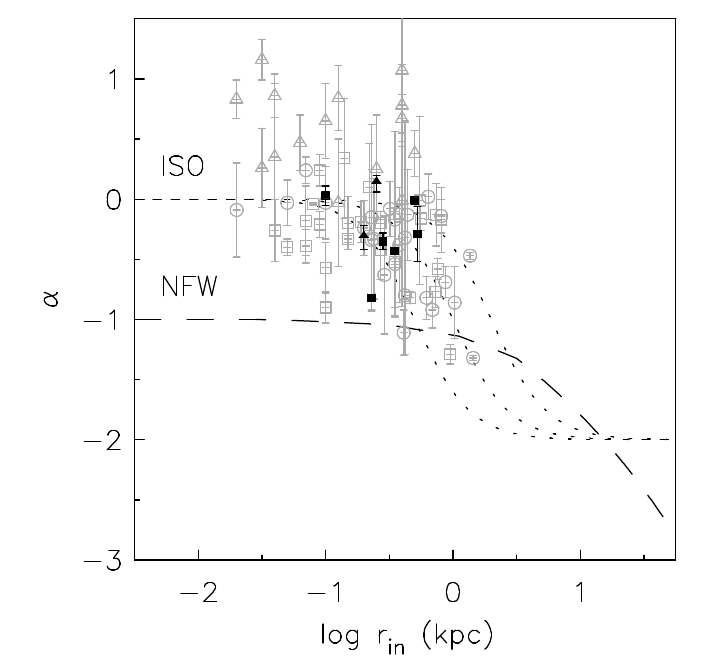}
\caption{The inner slope of the dark matter density profiles plotted against
  the radius of the innermost point. Presented in grey are the results from
  publications by \citet{deBlok2001a} (open circles), \citet{deBlok2002} (open
  squares), and by \citet{Swaters2003} (open triangles). Shown in black are the
  results from Oh08 (solid triangles). Our results are overplotted
  with black squares. We fit the isothermal and NFW profiles following
  \citet{deBlok2001a}.}
\label{alpha}
\end{figure}

We now take the plot by \citet{deBlok2001a} (see also Oh08),
which shows $\alpha$ \emph{vs.} the logarithm of the innermost radius (see
Fig.~\ref{alpha}). Data from previous studies are plotted with grey open
circles \citep{deBlok2001a}, grey open squares \citep{deBlok2002}, grey open
triangles \citep{Swaters2003}, and black solid triangles (Oh08). Our
own results are plotted with black solid squares and are in good agreement with
all previous works. Only IC\,5152, which has an inner slope of $\alpha=-0.82$,
is close to CDM predictions.

We now discuss possible systematic errors in the data (as mentioned in the
  introduction): \citet{Swaters2003} and \citet{Spekkens2005} used rotation
  curves of dwarf and low surface brightness galaxies obtained from optical
  long-slit spectra in order to measure the slopes of the mass
  distribution. They compared their results to model spectra by taking into
  account possible sources of uncertainties like slit misalignment, slit
  width, and seeing. Both studies came to the result that the galaxies do not
  require haloes with steep cusps, but that haloes with $\alpha=1$ cannot be
  ruled out. However, they clearly showed that a steeper slope ($\alpha>1$) is
  not consistent with the observed profiles. As we use 2d velocity fields, our
  data are not affected by the above mentioned errors. We can also rule out
  geometric effects caused by low velocity gas picked up along the line of
  sight at higher inclinations as our data have a very high spectral
  resolution.

As already mentioned in Sect.~\ref{SectN2366harm}, CT08 could show that
inaccurate centre positions do not change the inner slope of a rotation
curve. Non-circular motions are detected, but small enough to not
significantly change the rotation velocities. Nevertheless, there are problems
that affect our results as, e.g., the low spatial resolution of especially the
ATCA data.

Summarised it can be said that for the majority of galaxies studied here, the
ISO halo reproduces the observed rotation curves much better than the NFW
halo, which is in good agreement with previous studies
\citep[e.g.,][]{Kuzio2008,Spano2008}. The slopes of the observed rotation
curves are inconsistent with CDM predictions, but agree with the results from
  \citet{deBlok2001a}. One explanation for the cusp-core discrepancy might be
  that most simulations neglect the baryons. And indeed, first approaches to
  include baryons into CDM simulations show that their contribution changes
  the inner density profile from a cusp to a core \citep{Romano-Diaz2008}.
\begin{table} 
\caption{Slope $\alpha$ in the inner kpc.}
\label{tabalpha}
$$
\begin{tabular}{lccc}
  \hline
  \hline
  \noalign{\smallskip}
Galaxy & $\alpha$ & $\Delta \alpha$ & $r_{\rm in}$\\
& & & [kpc]\\
& (1) & (2) & (3)\\
\hline
\noalign{\smallskip}
NGC\,2366 & 0.03 & 0.08 & 0.10\\
ESO\,059-G001 & -0.29 & ... & 0.52\\
ESO\,215-G?009 & -0.35 & 0.23 & 0.28\\
NGC\,4861 & -0.01 & ... & 0.50\\
NGC\,5408 & -0.43 & ... & 0.35\\
IC\,5152 & -0.82 & 0.07 & 0.23\\
\noalign{\smallskip}
\hline
\end{tabular}
$$
\footnotesize{Notes: (1) the slope of the inner 1\,kpc of the density profile;
(2) uncertainties; (3) the radius of the innermost point.}
\end{table}
\section{Summary}
We used VLA and ATCA \HI\ synthesis data of a sample of six nearby irregular
dwarf galaxies in order to decompose the observed rotation curves into the
contributions from the stars, the gas, and the dark matter halo. In order to
rule out systematic effects, we first performed a harmonic decomposition of
the Hermite velocity fields. The results show that the
quadratically-added amplitude $A_{\rm r}(r)$ is generally below 6\skms, in
the central kpc even below 3\skms. Over the entire radial range non-circular
motions contribute less than 25\%, sometimes even less than 10\%\ to the local
rotation velocity (constrained case). According to \citet{Hayashi2004}, the
triaxial dark matter haloes cause non-circular motions as large as 50\%\ of
the local rotation velocity within the central kpc. Our much smaller
amplitudes demonstrate that the observed rotation curves of the galaxies
studied here are not severely influenced by non-circular motions.

With this knowledge we decomposed the observed rotation curves into their
components. We fitted a cuspy NFW halo and a cored pseudo-isothermal halo to
the rotation curves and modelled the minimum-disc, the minimum-disc + gas, and
the maximum-disc cases. The resulting model fits show that in all sample
galaxies (except for ESO\,215-G?009), the observed rotation curves are better
represented  by the ISO halo than by the NFW halo. As the outer parts of the rotation curves are affected
by uncertainties caused by the sparsely filled tilted rings, we also performed
a mass decomposition by fitting the inner few kpcs only. For most galaxies,
this significantly improved the quality of both the NFW and the ISO halo
models. The reduced $\chi^2$ could in most cases be decreased for both halo
models by including the gas. In some cases, a further improvement could be
achieved when also including the stars. This implies that the baryons should
not be neglected, at least not in the inner part of dwarf galaxies \citep[see
  also][]{Swaters2009}.

\begin{acknowledgements}
The authors would like to thank the referee for the useful comments that
significantly improved this paper.\\
Furthermore, we thank Nic Bonne and Erwin
de Blok for their help with the mass decomposition. We would also like to
thank Fabian Walter for providing the THINGS data of NGC\,2366 and Eric
Wilcots for providing the VLA data cubes of NGC\,4861.\\
This work was partly supported by the Deutsche Forschungsgesellschaft (DFG)
under the SFB 591, by the Research School of the Ruhr-Universit\"at Bochum,
and by the Australia Telescope National Facility, CSIRO. It is partly based on
observations with the Australia Telescope Compact Array. The ATCA is part of
the Australia Telescope which is funded by the Commonwealth of Australia for
operation as a National Facility managed by CSIRO. It is also partly based on
archival VLA data of the National Radio Astronomy Observatory. The NRAO is a
facility of the National Science Foundation operated under cooperative
agreement by Associated Universities, Inc.\\
We made use of NASA's Astrophysics Data System (ADS) Bibliographic Services
and the NASA/IPAC Extragalactic Database (NED) which is operated by the Jet
Propulsion Laboratory, California Institute of Technology, under contract with
the National Aeronautics and Space Administration.
\end{acknowledgements}
\bibliographystyle{aa}
\bibliography{Diss}
\end{document}